\documentclass[12pt]{article}
\usepackage{amsmath}
\usepackage{soul}
\usepackage{xcolor}
\usepackage{amssymb}
\usepackage{gensymb}

\usepackage{graphicx}
\usepackage[noblocks]{authblk}

\usepackage[utf8]{inputenc}
\usepackage[english]{babel}

\usepackage[sorting=none]{biblatex}
\makeatletter
\def\@maketitle{%
  \newpage
  \vspace{-8em}
  \begin{center}%
  \let \footnote \thanks
    {\huge \bfseries \@title \par}%
    \vskip 0.0em%
    {\normalsize
      \lineskip 0.em%
      \vskip 1.0em%
      LiquidO Collaboration\footnotemark
      \begin{tabular}{c}%
        \@author
      \end{tabular}\par}%
    \vskip 0.em%
    {\normalsize \@date}%
  \end{center}%
  \par
  }
\makeatother

\newcommand{\Londrina}{Departamento de F\'isica, 
	Universidade Estadual de Londrina, 
	Londrina,
	Brazil}

\newcommand{\PUCR}{Department of Physics, 
	Pontif\'icia Universidade Cat\'olica do Rio de Janeiro, 
	Rio de Janeiro, Brazil}

\newcommand{\Queens}{Department of Physics, 
	Engineering Physics \& Astronomy, 
	Queen's University, Kingston, 
	Canada}

\newcommand{\Prague}{Institute of Particle and Nuclear Physics,
	Charles University,
	Prague, 
	Czech Republic}

\newcommand{\IJCLabSaclay}{Universit\'e Paris-Saclay, 
	CNRS/IN2P3, 
	IJCLab, 
	Orsay, 
	France}

\newcommand{\LPtwoI}{Universit\'e de Bordeaux, 
	CNRS, 
	LP2I Bordeaux, 
	Gradignan, 
	France}

\newcommand{\SUBA}{Nantes Universit\'e, 
	IMT-Atlantique, 
	CNRS, 
	Subatech,
	Nantes, 
	France}

\newcommand{\FerraraUni}{Dipartimento di Fisica e Scienze della Terra, 
	Universit\`{a} di Ferrara, 
	Ferrara, 
	Italy}
\newcommand{\FerraraINFN}{INFN, 
	Sezione di Ferrara, 
	Ferrara, 
	Italy}

\newcommand{\Padova}{INFN, 
	Sezione di Padova, 
	Padova, Italy}

\newcommand{\MainzA}{Johannes Gutenberg-Universit\"{a}t Mainz,
	Institut f\"{u}r Physik, 
	Mainz, Germany} 
\newcommand{\MainzB}{Johannes Gutenberg-Universit\"{a}t Mainz,
	Detektorlabor, Exzellenzcluster PRISMA$^+$,
	Mainz, Germany} 	

\newcommand{\CIEMAT}{CIEMAT, 
	Centro de Investigaciones Energ\'{e}ticas, Medioambientales y Tecnol\'{o}gicas, 
	Madrid, Spain}

\newcommand{\UniZar}{Centro de Astropart\'{\i}culas y F\'{\i}sica de Altas Energ\'{\i}as (CAPA),
 	Universidad de Zaragoza, 
	Zaragoza, Spain}

\newcommand{\DIPC}{Donostia International Physics Center, 
	Basque Excellence Research Centre, 
	San Sebasti\'an/Donostia, 
	Spain}

\newcommand{\RCNS}{RCNS, 
	Tohoku University, 
	Sendai, Japan}

\newcommand{\Sussex}{Department of Physics and Astronomy, 
	University of Sussex, 
	Brighton, 
	United Kingdom}

\newcommand{\ImpCol}{Department of Chemistry, 
	Imperial College London, 
	London, 
	United Kingdom}
	
\newcommand{\RAL}{Rutherford Appleton Laboratory, 
	Didcot,
	Oxford,
	United Kingdom}

\newcommand{\UCI}{Department of Physics and Astronomy, 
	University of California at Irvine, 
	Irvine, 
	CA, 
	USA}

\newcommand{\UniPennPhys}{Department of Astronomy and Astrophysics, 
	Pennsylvania State University, 
	University Park, 
	PA,
	USA}
\newcommand{\UniPennAstro}{Department of Physics, 
	Pennsylvania State University, 
	University Park, 
	PA,
	USA}

\newcommand{\UniMichigan}{Department of Nuclear Engineering and Radiological Sciences,
 	University of Michigan, 
	Ann Arbor, 
	MI,
	USA}
					
\newcommand{\BNL}{Brookhaven National Laboratory, 
	Upton, 
	NY,
	USA}

\title{Characterization of a radiation detector based on opaque water-based liquid scintillator}

\begin{document}

%
%
\author[z]{J.\,Apilluelo}
\author[b]{L.\,Asquith}
\author[b]{E.\,F.\,Bannister.}
\author[p]{J.\,L.\,Beney}
\author[p]{X.\,de\,La\,Bernardie}
\author[b]{T.\,J.\,C.\,Bezerra}
\author[p]{M.\,Bongrand} 
\author[q$\alpha$]{C.\,Bourgeois}
\author[q$\alpha$]{H.\,Boutalha}
\author[q$\alpha$]{D.\,Breton}
\author[q$\alpha$]{M.\,Briere}
\author[q$\alpha$,c]{A.\,Cabrera} 
\author[p]{A.\,Cadiou}
\author[l]{E.\,Calvo}
\author[q$\alpha$]{V.\,Chaumat}
\author[f]{E.\,Chauveau}
\author[b]{B.\,J.\,Cattermole}
\author[h]{M.\,Chen}
\author[i]{P.\,Chimenti} 
\author[q$\alpha$]{T.\.Cornet}
\author[x$\alpha$,x$\beta$]{D.\,F.\,Cowen}
\author[q$\alpha$]{C.\,Delafosse}
\author[r$\alpha$]{S.\,Dusini} 
\author[b]{A.\,Earle}
\author[i]{C.\,Frigerio-Martins}
\author[z]{J.\,Gal\'an}
\author[q$\alpha$]{A.\,Gallas}
\author[z]{J.\,A.\,Garc\'ia}
\author[q$\alpha$]{R.\,Gazzini}
\author[b]{A.\,Gibson-Foster}
\author[m$\alpha$]{C.\,Girard-Carillo}
\author[b]{W.\,C.\,Griffith}
\author[u]{J.\,J.\,G\'omez-Cadenas}
\author[p]{M.\,Guitti\`ere}
\author[p]{F.\,Haddad}
\author[b]{J.\,Hartnell} 
\author[d]{A.\,Holin}
\author[q$\alpha$]{G.\,Hull} 
\author[z]{I.\,G.\,Irastorza}
\author[a]{I.\,Jovanovic}
\author[m$\alpha$]{L.\,Koch}
\author[q$\alpha$,c]{J.\,F.\,Le\,Du}
\author[h]{C.\,Lefebvre}
\author[p]{F.\,Lefevre}
\author[q$\alpha$]{F.\,Legrand}
\author[q$\alpha$]{P.\,Loaiza}
\author[b]{J.\,A.\,Lock}
\author[z]{G.\,Luz\'on}
\author[q$\alpha$]{J.\,Maalmi}
\author[j]{J.\,P.\,Malhado}
\author[e$\alpha$,e$\beta$]{F.\,Mantovani}
\author[f]{C.\,Marquet} 
\author[z]{M.\,Mart\'inez}
\author[q$\alpha$]{B.\,Mathon}
\author[q$\alpha$,l]{D.\,Navas-Nicol\'as}
\author[t]{H.\,Nunokawa} 
\author[g]{J.\,P.\,Ochoa-Ricoux} 
\author[l]{C.\,Palomares} 
\author[d]{D.\,Petyt}
\author[p]{P.\,Pillot}
\author[b]{J.\,C.\,C.\,Porter} 
\author[f]{M.\,S.\,Pravikoff}
\author[q$\alpha$]{H.\,Ramarijaona}
\author[f]{M.\,Roche}
\author[y]{R.\,Rosero}
\author[q$\alpha$]{P.\,Rosier}
\author[s]{B.\,Roskovec}
\author[z]{M.\,L.\,Sarsa}
\author[m$\beta$]{S.\,Schoppmann}
\author[r$\alpha$]{A.\,Serafini} 
\author[d]{C.\,Shepherd-Themistocleous}
\author[b]{W.\,Shorrock}
\author[q$\alpha$]{L.\,Simard}
\author[u]{S.\,R.\,Soleti}
\author[m$\alpha$,m$\beta$]{H.\,Th.\,J.\,Steiger}
\author[p]{D.\,Stocco}
\author[e$\alpha$,e$\beta$]{V.\,Strati}
\author[p]{J.\,S.\,Stutzmann}
\author[v]{F.\,Suekane} 
\author[m$\alpha$]{A.\, Tunc}
\author[b]{N.\,Tuccori}
\author[l]{A.\,Verdugo}
\author[p]{B.\,Viaud}
\author[m$\alpha$]{S.\,M.\,Wakely}
\author[m$\alpha$]{A.\,Weber}
\author[x$\beta$]{G.\,Wendel}
\author[a,*]{A.\,S.\,Wilhelm}
\author[y]{M.\,Yeh}
\author[p]{F.\,Yermia}
%
%
%
%
%
%
\affil[a]{\UniMichigan} 
\affil[b]{\Sussex} 
\affil[d]{\RAL} 
\affil[e$\alpha$]{\FerraraINFN} 
\affil[e$\beta$]{\FerraraUni} 
\affil[f]{\LPtwoI} 
\affil[g]{\UCI} 
\affil[h]{\Queens} 
\affil[i]{\Londrina} 
\affil[j]{\ImpCol} 
\affil[l]{\CIEMAT} 
\affil[m$\alpha$]{\MainzA} 
\affil[m$\beta$]{\MainzB} 
\affil[p]{\SUBA} 
\affil[q$\alpha$]{\IJCLabSaclay} 
\affil[r$\alpha$]{\Padova} 
\affil[s]{\Prague} 
\affil[t]{\PUCR} 
\affil[u]{\DIPC} 
\affil[v]{\RCNS} 
\affil[x$\alpha$]{\UniPennPhys} 
\affil[x$\beta$]{\UniPennAstro} 
\affil[y]{\BNL} 
\affil[z]{\UniZar} 

\affil[*]{Corresponding author: Andrew Wilhelm, andhelm@umich.edu or LiquidO-Contact-L@in2p3.fr}

\maketitle

\newpage


\begin{abstract}
We present the characterization of a novel radiation detector based on an opaque water-based liquid scintillator. Opaque scintillators, also known as LiquidO, are made to be highly scattering, such that the scintillation light is effectively confined, and read out through wavelength-shifting fibers.  The 1-liter, 32-channel prototype demonstrates the capability for both spectroscopy and topological reconstruction of point-like events.  The design, construction, and evaluation of the detector are described, including modeling of the scintillation liquid optical properties and the detector's response to gamma rays of several energies.  A mean position reconstruction error of 4.4~mm for 1.6~MeV-equivalent events and 7.4~mm for 0.8~MeV-equivalent events is demonstrated using a simple reconstruction approach analogous to center-of-mass.
\end{abstract}


\section{Introduction}
\label{sec:Introduction}

Most liquid scintillator radiation detectors collect the light generated by an interacting particle and direct it to a photosensor such as a photomultiplier tube (PMT) or a silicon photomultiplier (SiPM) at the scintillator boundary \cite{Alimonti2000LightScintillator,Ito1995ADetector}.  To maximize the light collection, scintillation materials are chosen to be as transparent as possible to their fluorescence.  Scintillation light is emitted isotropically, so much of the light collected at the photosensor may arrive after reflection or scatter at non-sensing boundaries, such as detector walls.  The photosensors used are generally not position-sensitive, and even for position-sensitive photosensors, the correlation between the distribution of detected light and its vertex is complicated by absorption, scattering, and reflections in the detector volume.  For these reasons, detectors of this type generally do not provide high-fidelity information on where in the detector volume an event occurred.  

The two main categories of methods for obtaining event topological information in a traditional scintillation detector are imaging the interior of the detector, usually by using multiple photosensors spread out around the outside of the detector volume~\cite{Suzuki2019TheExperiment, Li2021KamLAND-ZenAnalysis, Wagner2018AMeasurements, Askins2020Theia:Detector, Harvey2021DevelopmentApplications,Manfredi2020TheCamera,Ranucci2001Borexino,Chen2008TheExperiment,DoubleChooz,An2016NeutrinoJUNO}, and physically segmenting the detector into voxels~\cite{Li2019AArrays,Sutanto2021SANDD:Scintillator,Weinfurther2015LowCamera,Manfredi2020TheCamera,Ayres2007TheReport,Aliaga2014DesignDetector, Michael2008TheExperiment, Abreu2021SoLid:Experiment,Ashenfelter2019TheExperiment,Abe2011TheExperiment,Blondel2018AViews}.

By contrast, opaque scintillators \cite{Buck2019,Cabrera2024OpaqueScintillation,Schoppmann2022ReviewPhysics,Tang2020ProspectsExperiments}, which are designed to have a short scattering length, seek to constrain the scintillation photons by repeated scattering.  The light produced by an event is \textit{stochastically confined} about its origin, and information about the event topology is thus retained~\cite{Cabrera2021, Cabrera_2022_Talk}.  The scintillation photons are collected via a lattice of wavelength-shifting (WLS) fibers, each of which is read out into a separate data acquisition channel, such that the number of hits in each fiber gives information about where in the volume the light was collected.  This information can then be used to reconstruct the event topology and energy.

This pioneering technology, known as LiquidO, is currently being developed by the international LiquidO Collaboration~\cite{LiquidO_webpage}.
The LiquidO concept was originally conceived to enhance particle-type identification and increase the signal-to-noise ratio in large antineutrino detectors through analysis of event topology.  Two examples of such projects are CLOUD \cite{Hartnell2023ChoozDetector} and Super Chooz \cite{Cabrera2019PossibleUnitarity}.  However, it is also being considered for positron emission tomography (PET) \cite{LPET_Conference_Presentation_2022}, improving energy resolution for photons in calorimeters for use with high-energy e$^+$e$^-$ colliders~\cite{Hull2021ProofCalorimeter}, and the detection of geoneutrinos~\cite{Cabrera2023ProbingGeoneutrinos}.  

This article describes the characterization of a prototype LiquidO radiation detector capable of providing spectroscopic and vertex information.  Such a detector could potentially be used for applications currently served by traditional optically segmented detectors, such as gamma-ray and neutron-scatter cameras, or for muon scatter tomography.  The performance of these types of detector systems is generally dependent on the precision of vertex reconstruction, so an improvement in position resolution as can be gained in an opaque scintillator is desirable. Specifically, the detector gamma-ray response for the prototype is shown, and the reconstruction quality for point-like events is quantified.

Section~\ref{sec:ApparatusAndMaterial} describes the experimental apparatus and materials, including the detector design, the data acquisition scheme, and the various scintillation liquids investigated.  While other categories of scintillators have been considered for opaque scintillation detectors, including wax-based (NoWaSH-20~\cite{Buck2019}) and liquid scintillator cooled to liquid xenon temperatures~\cite{PandaX}, in this study we introduce opaque emulsion-based liquid scintillators based on the principle of water-based liquid scintillator.  Section~\ref{sec:Results} contains the experimental results, including the measurement of light yields of each of the liquid scintillators and the determination of optical parameters for one of them.  Then, response to gamma rays is shown for each scintillation liquid and the vertex reconstruction capability of one of them is demonstrated using pulses of photons from a fiber-coupled laser.  Lastly, in Sec.~\ref{sec:Conclusion}, we give context for the achieved precision of vertex reconstruction and discuss planned future work. 
\section{Apparatus and material}
\label{sec:ApparatusAndMaterial}
The efficiency of light collection, transport, and conversion to signal for an experimental apparatus that shares most components with the one in this work was characterized in Ref.~\cite{Wilhelm2023EvaluationFibers}.  The type of fiber (Kuraray Y-11(200)), 64-channel multianode photomultiplier tube (MAPMT) (Hamamatsu H12700A), and digitizer (CAEN V1730) were the same for this experiment as in that previous work.  An opaque formulation of water-based liquid scintillator (discussed in Sec.~\ref{sec:oWbLS}) was chosen due to its promising optical characteristics and material compatibility, and is referred to as opaque Water-based Liquid Scintillator (oWbLS).\footnote{Water-based liquid scintillator (WbLS) is often, but not always, majority water by volume.  This term clarifies that water is added to a traditional scintillator (oil-based solvent) via stable emulsion to create a different type of scintillator.  The formulation described in this experiment includes a small fraction of water to promote scattering, and the majority of its volume is an oil-based solvent.}
\subsection{Detector design and construction}
\label{sec:detector_design}
\begin{figure*}
    \centering
    \includegraphics[width=0.9\textwidth]{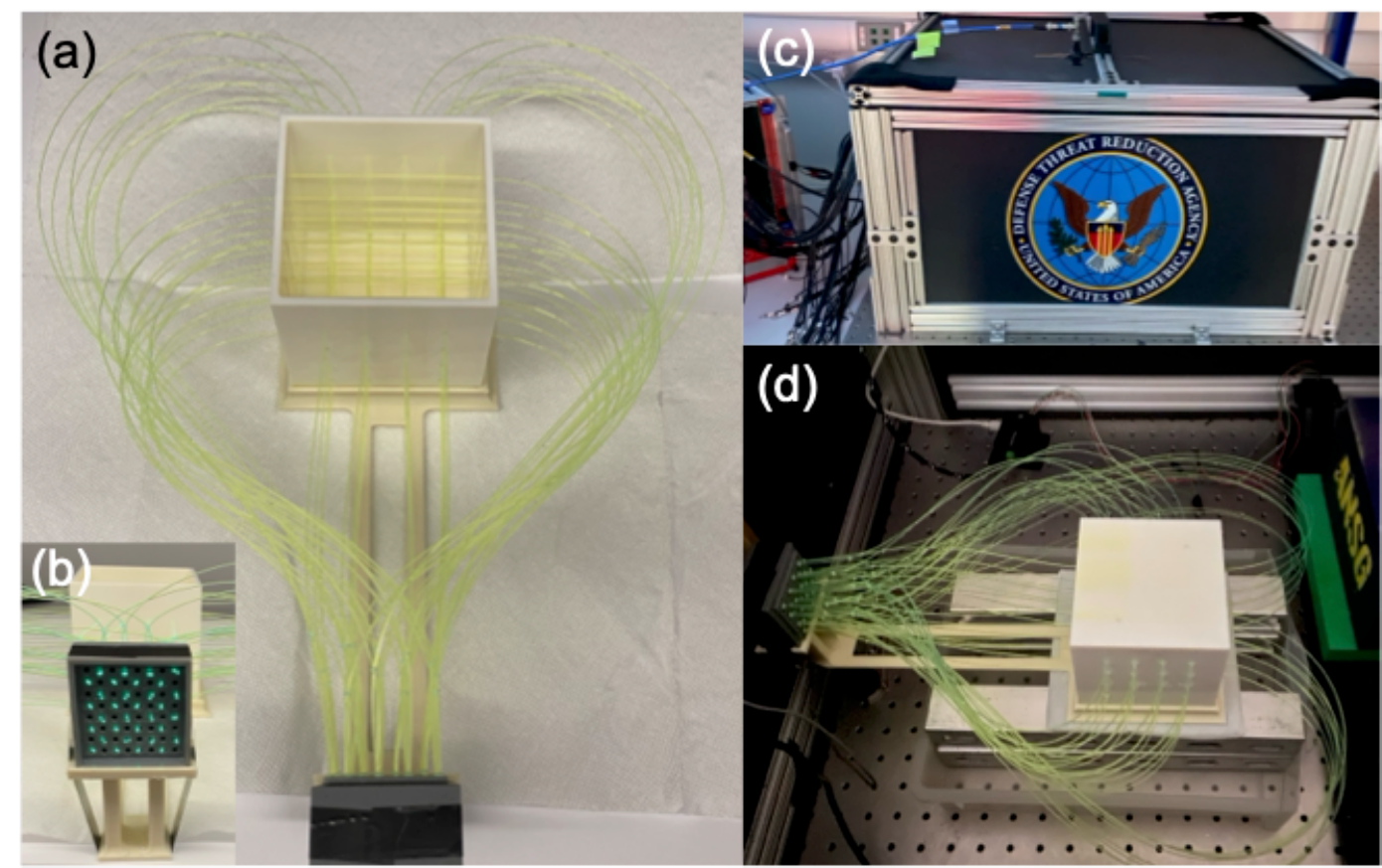}
    \caption{(a)~Detector body made of VeroWhite with 32 WLS fibers glued in place.  The fibers terminate at the MAPMT interface, shown from the front in~(b). (c)~Exterior of the dark box, with a pulsed LED mounted on top for calibration.  The readout electronics are on the left. (d)~Interior of the dark box with the detector mounted to the MAPMT. The top of the detector body is closed with a lid, also made of VeroWhite.}
    \label{fig:apparatus}
\end{figure*}

\begin{figure*}
    \centering
    \includegraphics[width=0.95\textwidth]{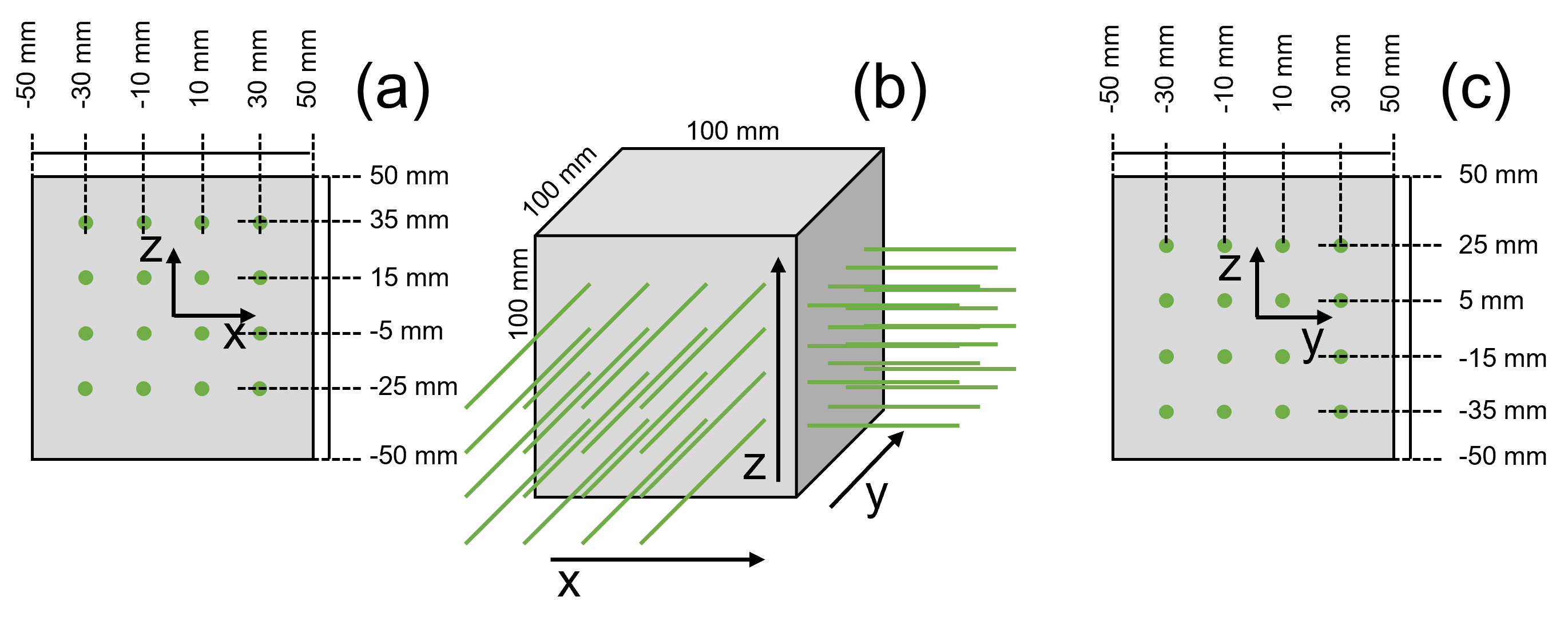}
    \caption{A sketch of the detector geometry.  (a)~The layout of the fibers in the $xz$~plane, (b)~a three-dimensional view of the detector body, and (c)~the layout of the fibers in the $yz$~plane.  The origin of the coordinate system is the center of the detector volume.}
    \label{fig:det_sketch}
\end{figure*}

Figure~\ref{fig:apparatus} shows the experimental apparatus. The body of the detector was made from an additively manufactured material (VeroWhite).  The wall reflectivity is discussed in Sec.~\ref{sec:oWbLS_optical_parameter_determination}.  Due to the inherent light confinement of the design and the establishment of a fiducial region away from the walls, reflected light plays an insignificant role in the reconstructions of point-like events (further discussed in Sec.~\ref{sec:reconstruction}).  The detector volume was a 1~L cube, 10~cm on a side.  There were 16 WLS fibers arranged in a grid in each vertical plane.  The fibers in the $xz$ plane were offset 1~cm higher than the $yz$ plane to avoid collision and provide additional information for the $z$ direction.  Figure~\ref{fig:det_sketch} shows a sketch of the detector geometry.  The fibers were glued in place with a fast-curing epoxy (5-minute Z-POXY), which was found to be compatible with WLS fibers and to cause less light loss than other glues tested~\cite{Bannister2023InvestigatingFibres}.

The fiber pitch in the $x$- and $y$-directions was 2~cm, and 1~cm in the $z$-direction. 
Each fiber was approximately 1~m long.  The detector was designed and constructed before receiving any oWbLS samples, so the fiber pitch was not optimized for the optical properties of a particular formulation of oWbLS.  The length of the fibers was chosen to ensure that the minimum bending radius met the manufacturer's recommendations for minimizing light loss.
The detector, fibers, and MAPMT were housed in a dark box to minimize background light and to protect light-sensitive components.
\subsection{WLS fibers and MAPMT}
The WLS fibers were Kuraray Y-11(200) (non S-type) blue-to-green shifters, 1~mm in diameter~\cite{2022KurarayDatasheet}.  The fibers were cut with a precision cleaver and then polished with fine-grit sandpaper.  Both ends of each fiber terminated at a single pixel of the Hamamatsu H12700A MAPMT and were held in place by a printed interface and silicone optical grease.  The pixel size of the MAPMT is 36~mm$^2$, so the two fiber ends fit comfortably in the center of the pixel.  The channels were chosen in a checkerboard pattern on the MAPMT to minimize crosstalk, as shown in Fig.~\ref{fig:apparatus}(b).  The quantum efficiency of the MAPMT peaks at $\sim$32\% at 330~nm, but is $\sim$16\% at 500~nm, which is the approximate emission wavelength of Y-11 fibers~\cite{ 2022KurarayDatasheet,2019HamamatsuDatasheet}.
\subsection{Opaque water-based liquid scintillator}
\label{sec:oWbLS}
Water-based liquid scintillators have been under development at Brookhaven National Laboratory since 2011~\cite{Yeh2011AApplications}.  They were originally conceived as a method to tune the relative levels of scintillation and Cherenkov light so that both could be readily detected in large antineutrino detectors.  A stable emulsion is produced by introducing a surfactant that encapsulates the water into micelles.  Transparent WbLS is manufactured in such a way as to minimize scattering.  For this application, in which measuring the Cherenkov light separately from scintillation light is not an objective, the loading of water was minimized to maximize total light yield.  In a parallel development, the NoWaSH scintillator demonstrated a significant enhancement in light yield, producing about 80 times more light than pure water when exposed to 1.8 MeV electrons in the MINI prototype \cite{Cabrera_2022_Talk}.  For both scintillator types, the goal is to maximize the light yield by minimizing material that does not contribute to scintillation.

In traditional WbLS, water is used to reduce the light yield, whereas in oWbLS the water introduces opacity.  The micelles constitute scattering centers, and the appearance of the liquid becomes like soapy water.  The organic component of the WbLS used in this study is a mix of modified polyethylene glycol-based surfactants in $>$90\% of di-isopropylnaphthalene (DIN) liquid scintillator loaded with a fluor of 2,5-diphenyloxazole (PPO) at 3~g/L.  We made three oWbLS formulations, the difference between which is the water content in the organic mixture. Different water content induced different opacities (scattering lengths) in the oWbLS. In this study, 1.5\% of water was introduced in oWbLS1 while 2\% and 2.5\% water were added to the oWbLS2 and oWbLS3, respectively, to study detector response with different liquid scattering lengths.  In addition, the wavelength shifter 1,4-Bis(2-methylstyryl)benzene (bis-MSB) was included at 15mg/L to match the emission spectrum with the absorption spectrum of the WLS fibers.

\begin{figure}
    \centering
    \includegraphics[width = .75\textwidth]{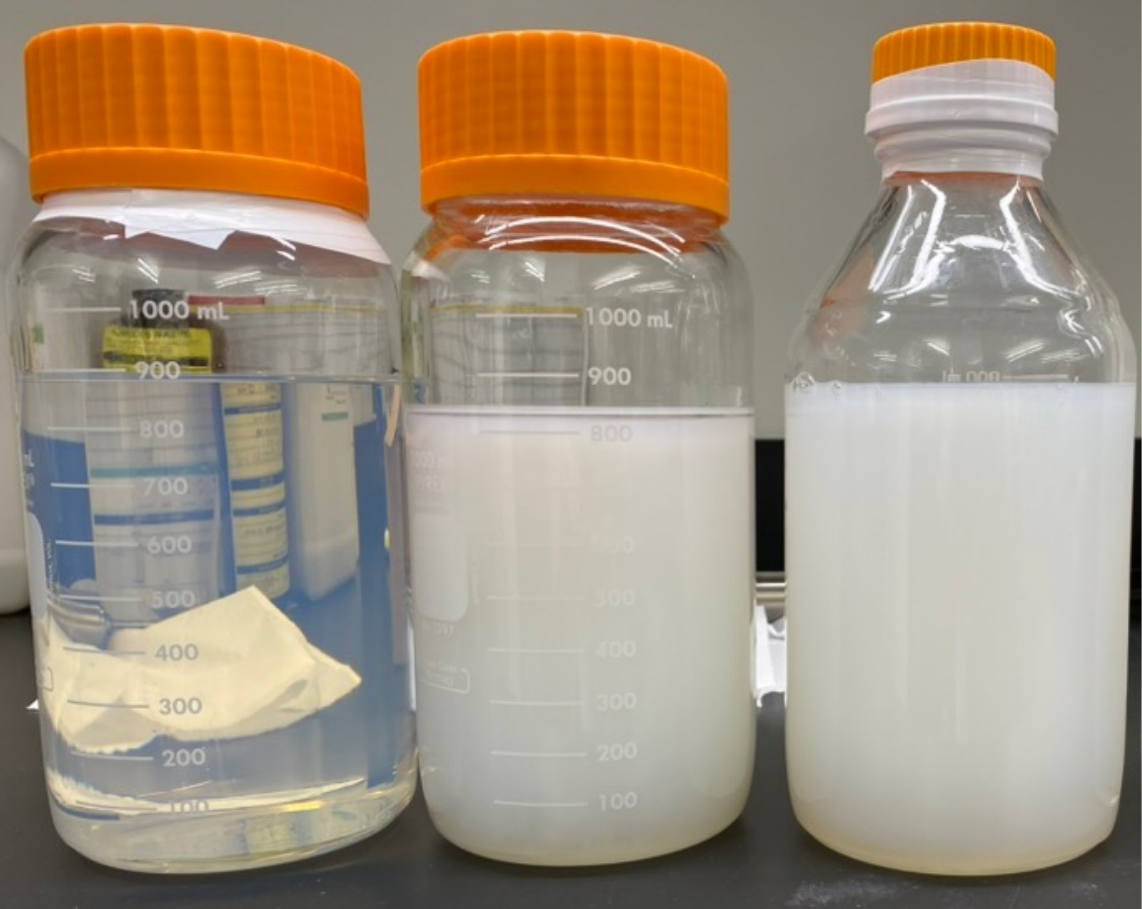}
    \caption{From left to right: oWbLS1, oWbLS2, oWbLS3.  oWbLS1 has the longest scattering length of the three oWbLS samples and appears relatively transparent, and oWbLS2 has the shortest scattering length.}
    \label{fig:scintillators}
\end{figure}

Images of the three formulations of oWbLS (oWbLS 1--3) are shown in Fig.~\ref{fig:scintillators}.

The initial batch of oWbLS2 showed some separation after shipment but was re-stabilized by bubbling nitrogen through the sample for 1~hour.  This separation may have been caused by exposure to heat during transport.  A lab-retained sample of the same batch did not show separation.  The sample of oWbLS3 became gradually less opaque over a period of a month, potentially due to the agglomeration of micelles.  A second batch of oWbLS2, used for the event reconstruction portion of this work, has shown good stability over approximately two months of data taking.

The scintillator was stored in a temperature-controlled room at approximately $20\degree$~C.  It was frequently exposed to the air throughout the experimental campaign.  While quenching due to oxygenation in a similar scintillator has been shown to be on the order of 10\%~\cite{Ashenfelter2018PerformanceExperiment}, such a measurement has not yet been made for oWbLS.
\subsection{Data acquisition and analysis}
\begin{figure}
    \centering
    \includegraphics[width=0.85\textwidth]{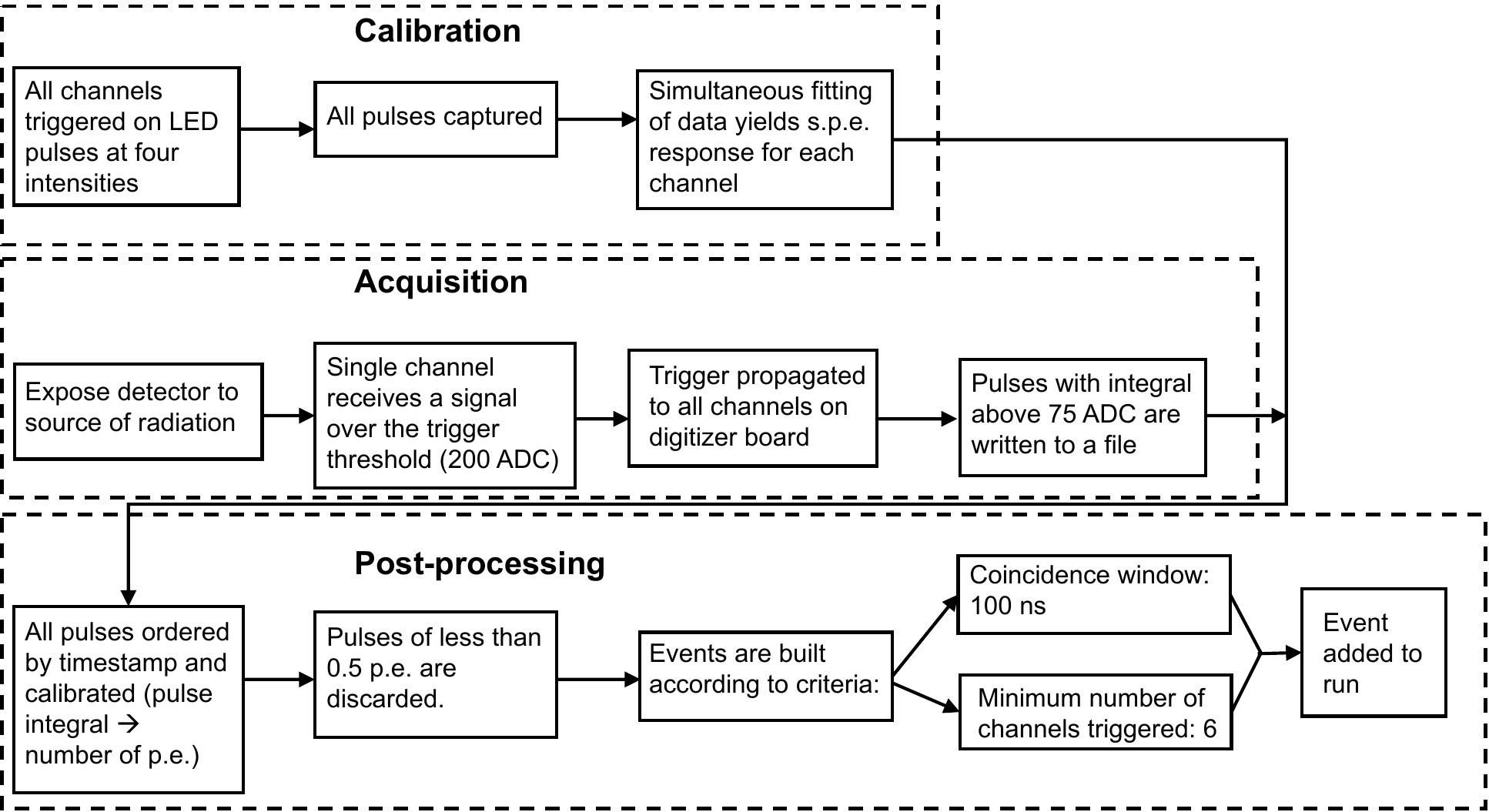}
    \caption{A flowchart of the data acquisition and analysis process.  The primary input for a data run is the s.p.e response for each channel, and the output is a list of events, each of which contains pulses from multiple channels that have been correlated by their timestamps and calibrated to be in units of p.e.}
    \label{fig:data_flowchart}
\end{figure}

Figure~\ref{fig:data_flowchart} shows the data acquisition and event-building process as a flowchart. 
Each channel of the MAPMT was read out into a channel of a 16-channel CAEN V1730 digitizer.  Two digitizers were used for a total of 32 data acquisition channels.  The response of each channel of the MAPMT was calibrated using a pulsed light-emitting diode (LED) mounted on the top of the dark box.  The single photoelectron (s.p.e) response of each channel was found using the procedure outlined in Ref.~\cite{Wilhelm2023EvaluationFibers}.  During an experimental run, the clocks of the two digitizers were synchronized to facilitate the compilation of events using event timestamps.  A voltage pulse that met the trigger threshold in a single channel resulted in the propagation of a trigger signal to all channels on that board, but only pulses that met a pulse-integral threshold of approximately 0.5~photoelectrons were kept.  The pulse-integral window was 100~ns, and integration was carried out onboard by the CAEN DPP-PSD firmware.  In post-processing, events were constructed by identifying pulses that fell within a 100~ns coincidence window.  This was necessary to correlate events between boards.  The output of the data analysis pipeline was a list of events, for which the number of photoelectrons generated in each channel per event was known.  This method of triggering and filtering was likely superfluous for this detector system because the background rate was very low.  However, in future detector designs that use silicon photomultipliers, this style of event selection will be required to reduce the dark count rate, so it was implemented here.  
\section{Results}
\label{sec:Results}
This section first describes the determination of optical parameters of the opaque scintillation liquids and then shows the response of each gamma-ray source.  These measurements are used to validate a simulation model for one of the liquids, which is in turn used to create a mapping of true and reconstructed event vertex positions.  The vertex reconstruction for events created by the injection of laser light into the detector is demonstrated and the mean reconstructed error is quantified.
\subsection{oWbLS light yield}
\label{sec:light_yield}
The light yield of oWbLS was measured using the method described in Ref.~\cite{Cumming2019ImprovingScintillators}, in which two large photomultiplier tubes view a cuvette of liquid scintillator exposed to gamma rays from $^{137}$Cs. 
The apparatus was a Beckman LS-6500 liquid scintillation counter.  The gamma rays from $^{137}$Cs interact with the scintillator samples primarily through Compton scattering, which results in a distribution of pulse heights that includes a discernible Compton edge.  The technique described in Ref.~\cite{Cumming2019ImprovingScintillators} is a way to extract the location of the Compton edge.  It involves first converting the light output spectrum to a power spectrum, fitting the power spectrum with a function, and then taking the derivative of that function.  The location of the Compton edge is taken to be the maximum of the derivative of the fitted function.  The light yield of an unknown scintillator is determined by comparing the location of the Compton edge of the unknown sample to a characterized reference.  The reference scintillator was linear alkylbenzene (LAB) with 3~g/L of PPO, which has a light yield between 8700~\cite{Buck2019ProductionExperiment} and 11100~\cite{Caravaca2020CharacterizationSeparation} photons/MeV.  The average of these two values was taken as the reference value for the light yield of LAB+PPO.
The light yield of oWbLS was determined to be 12000$\pm$2000 photons/MeV.  This value is consistent with other scintillators based on DIN, such as EJ-309,  which has a light yield of 12300 photons/MeV~\cite{EJ-309Technology}.  The uncertainty is dominated by the range of light yields reported in the literature for the reference scintillator.  These results are summarized in Table~\ref{table:light_yield}.

\begin{table}[ht]
\centering
\begin{tabular}{c p{0.9in} p{0.8in} } 
\hline\hline
 Material   & Measured light yield   & Relative light yield\\ 
 \hline
 LAB+PPO    & 10000$\pm$2000          &  1       \\ 
 oWbLS1     & 11000$\pm$2000         &  1.1       \\ 
 oWbLS2     & 12000$\pm$2000         &  1.2       \\
 oWbLS3     & 12000$\pm$2000         &  1.2       \\ 
 \hline \hline
\end{tabular}
\caption{Measured light yields of opaque scintillation liquids used in experiments.  oWbLS is composed primarily of DIN and has only a small fraction of water (less than 10\%), which explains its high light yield.}
\label{table:light_yield}
\end{table}

Although oWbLS has a relatively high light yield, the photon population in a LiquidO detector is attritted by several processes before it reaches the photosensor.  Many of these steps are not optimized in this prototype, and improvements in the efficiencies of each will be the focus of future work.

First, not all photons generated in this prototype will reach a WLS fiber.  This could be improved in future designs by increasing the absorption length of the scintillator and by optimizing the fiber pitch for a given set of scintillator optical properties.  Second, only some of the photons that reach a fiber are absorbed and re-emitted.  This could be optimized through better matching the WLS absorption spectrum to the scintillator emission spectrum.  Third, most wavelength-shifted photons are re-emitted in a direction that does not result in total internal reflection and thus are not transported to the photosensor.  Instead, they immediately escape the fiber.  This causes the greatest loss of any of the steps described here.  Kuraray estimates that their multi-clad fibers have a trapping efficiency of at least 10.8\% for double-end fiber readout~\cite{2022KurarayDatasheet}, meaning that almost 90\% of photons emitted in the WLS fiber core are lost.  These photons have already been wavelength-shifted, and therefore cannot be reabsorbed by another WLS fiber.  Investigation into the improvement of the effective fiber trapping efficiency is ongoing.  Fourth, some photons that are transported in the fiber core are absorbed before reaching the photosensor.  Because all the fibers in this design terminate at a single MAPMT, the fibers are much longer than they would need to be if the photosensors were located immediately outside the active volume.  Last, less than half of the photons that reach the photosensor will generate a measurable electric signal, a probability described as the device quantum efficiency.  This loss is not unique to LiqudO detectors but is unavoidable in any detector that uses light to detect radiation.  However, the quantum efficiency of the MAPMT used in this prototype is less than a typical SiPM, so future designs have the opportunity for improvement in this step as well.

Overall, while stochastic photon confinement makes photon collection a more complicated process than in a transparent scintillator, we expect future prototypes to improve on the efficiency of this initial design.

\subsection{oWbLS absorption and scattering length}
\label{sec:oWbLS_optical_parameter_determination}
The degree of light confinement in an opaque scintillator depends on its scattering and absorption lengths, as well as the scattering anisotropy.  In this analysis, we account for the effect of scattering anisotropy by using the reduced scattering coefficient
\begin{align}
\mu'_s = \mu_s(1-g),
\end{align}
in which $\mu_s$ is the true scattering coefficient and $g$ is an anisotropy parameter between 0 and 1.  The reduced scattering length is therefore given by
\begin{align}
    \lambda' = 1/\mu'_s.
\end{align}

The Geant4 model uses Rayleigh scattering, which is isotropic~\cite{Saidi1995MieSkin}.  In reality, the scattering may be in the Mie regime, meaning that the scattering center size is comparable to light wavelength, and is thus anisotropic~\cite{Bunge2006RayleighFibers}.  The use of reduced scattering length in the model accounts for the potential anisotropy of Mie scattering in the experiment.

\begin{figure}
    \centering
    \includegraphics[width=0.95\textwidth]{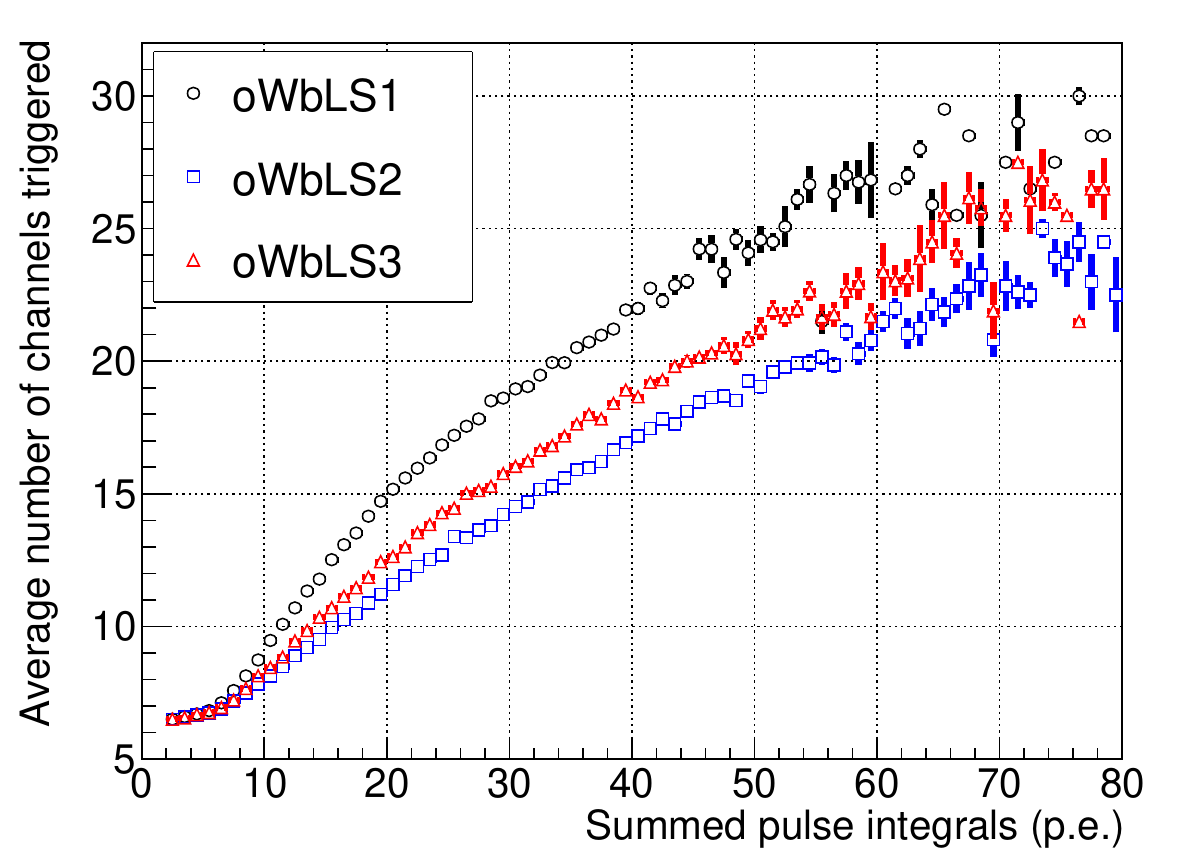}
    \caption{Relationship between the number of triggered channels and total event signal for the three formulations of oWbLS.  The data shown was generated by exposing the detector to gamma rays from a $^{60}$Co source placed on the center of the detector lid.  $^{60}$Co emits two gamma rays, at 1173.2~keV and 1332.5~keV.  The maximum energy deposition from a single scatter of the 1332.5~keV gamma ray is 1118~keV.  A liquid scintillator with an extremely short scattering length (\textit{e.g.} 1\% of the fiber pitch) would produce events in which only one or two channels were triggered, regardless of the total intensity of the event.  The more channels triggered per event for a given summed pulse integral, the less confining the liquid is to the scintillation photons.}
    \label{fig:light_confinement_comparison}
\end{figure}

An increase in either the reduced scattering or absorption length increases the mean distance a photon travels from its origin before being absorbed, and thus increases the size of the \textit{light ball}~\cite{Cabrera2021}.  One way to compare the relative size of the light ball among different oWbLS formulations is to observe the number of channels triggered as a function of the total signal collected in an event.  For a highly confining formulation, the number of channels increases slowly as the total signal increases, whereas, for a transparent scintillator, the number of channels triggered increases more quickly.  This relationship is apparent in Fig.~\ref{fig:light_confinement_comparison}.

\begin{figure}
    \centering
    \includegraphics[width=0.95\textwidth]{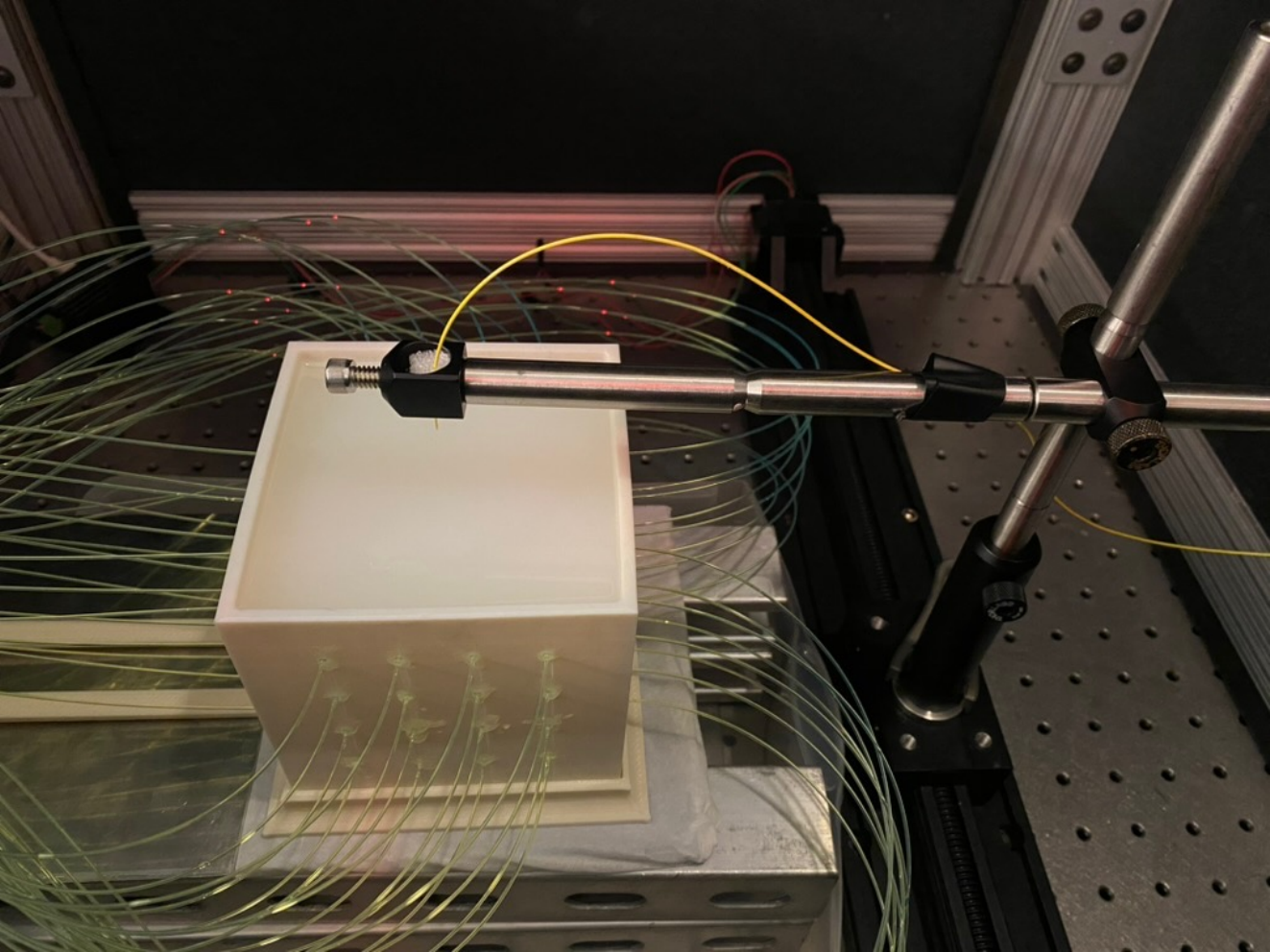}
    \caption{An image of the measurement described in Sec.~\ref{sec:oWbLS_optical_parameter_determination}.  A fiber-coupled laser is used to inject photons into the center of the detector, and the response for each channel is measured.  The laser fiber is held by an apparatus mounted on a linear translation stage so that the measurement can be repeated at various distances from the center of the detector.}
    \label{fig:laser_measurement}
\end{figure}

To determine the absorption and reduced scattering length of oWbLS2, a pulsed fiber-coupled laser was used to inject a known number of photons into the detector volume, as shown in Fig.~\ref{fig:laser_measurement}.  
The measured distribution of signal over WLS fiber channels was compared to a Geant4 simulation.  The model includes tracking each photon until it is ultimately absorbed, escapes the detector volume, or reaches the photocathode of the MAPMT.  When a photon reaches the photocathode of the MAPMT, an application of the wavelength-dependent quantum efficiency of the MAPMT dictates whether it is counted as a hit.  The model of WLS fiber includes both layers of the cladding and the wavelength-shifting process that takes place in the core.  The parameters used to model fibers, including their absorption and emission spectra, are described in detail in Ref.~\cite{Wilhelm2023EvaluationFibers}.  Six parameters were allowed to vary: $x$, $y$, and $z$ position of the fiber tip, absorption and reduced scattering length of the oWbLS, and the reflectivity of the detector walls.  The photons in the simulation were given an initial momentum in the $-z$ direction to model the photons emitted from the laser fiber.  By optimizing these parameters to minimize the residual between simulated and experimental data, estimates of the parameters can be made.  The estimated position of the fiber tip before fitting the data was ($0\pm5$(x),$0\pm5$(y),$0\pm5$(z))~mm, with the center of the detector volume taken as the origin.      

The wall reflectivity was estimated from Ref.~\cite{Golhin2023AppearanceJetting}, which quantified the optical parameters of several colors of printed plastics, to be approximately 0.3.  The wavelength of the laser ($407$~nm) lies in the transition region for the spectral reflectance of VeroWhite plastics.  The indices of refraction of liquid scintillation cocktails based on DIN were found to be between 1.55 and 1.6 at 404.7~nm, with most around 1.56~\cite{Kossert2013MeasurementCocktails}.  The index of refraction of oWbLS is likely to be slightly smaller as a result of the addition of water.  The index of refraction of VeroClear, a transparent but similar material to VeroWhite, was reported to be about 1.48 at 400~nm~\cite{Song2015Matching-index-of-refractionVisualization}, so the specular reflection at the oWbLS-VeroWhite boundary is expected to be less than in the measurement of Ref.~\cite{Golhin2023AppearanceJetting}, which was done in air.

\begin{figure*}
    \centering
    \includegraphics[width=\textwidth]{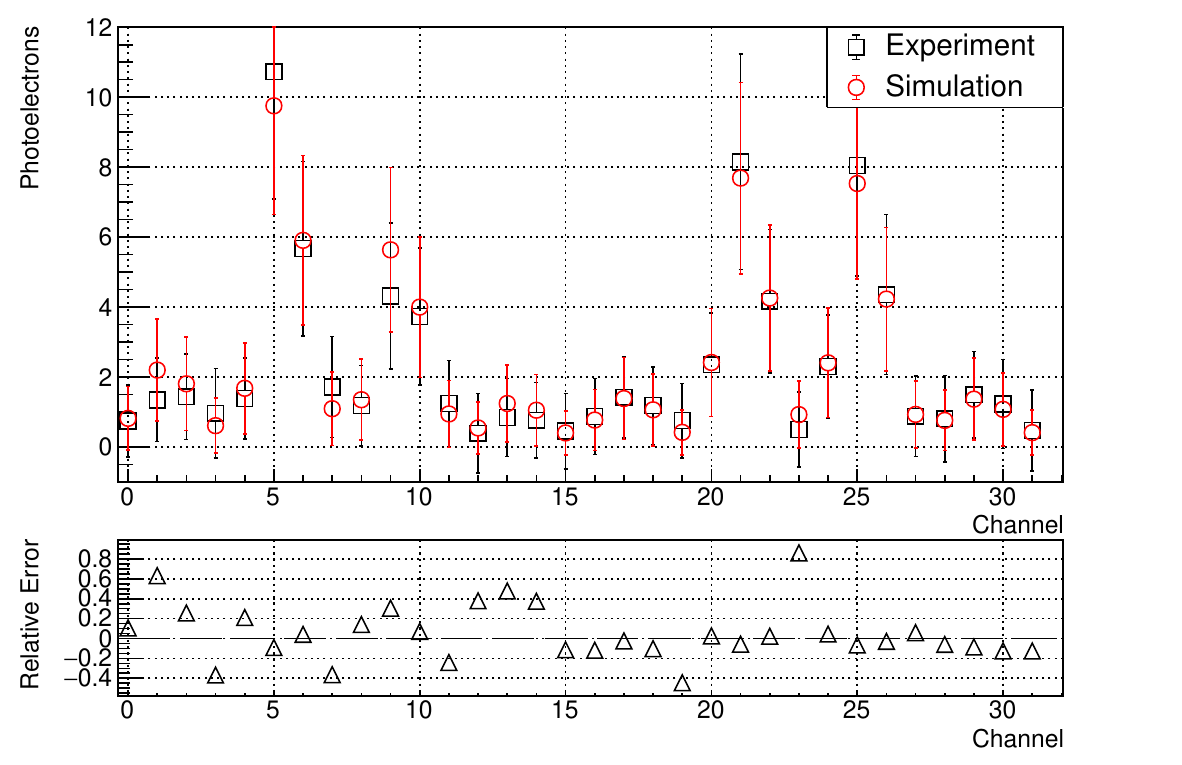}
    \caption{A comparison of the average signal in each channel for laser pulses with $20000\pm2000$ photons injected near the center of the detector.  The optical parameters used in the simulation are the values found by the genetic algorithm.  The vertical error bars show the standard deviation of the signal in each channel across 10000 pulses.}
    \label{fig:parameter_search_results}
\end{figure*}

\begin{figure*}
    \centering
    \includegraphics[width=\linewidth]{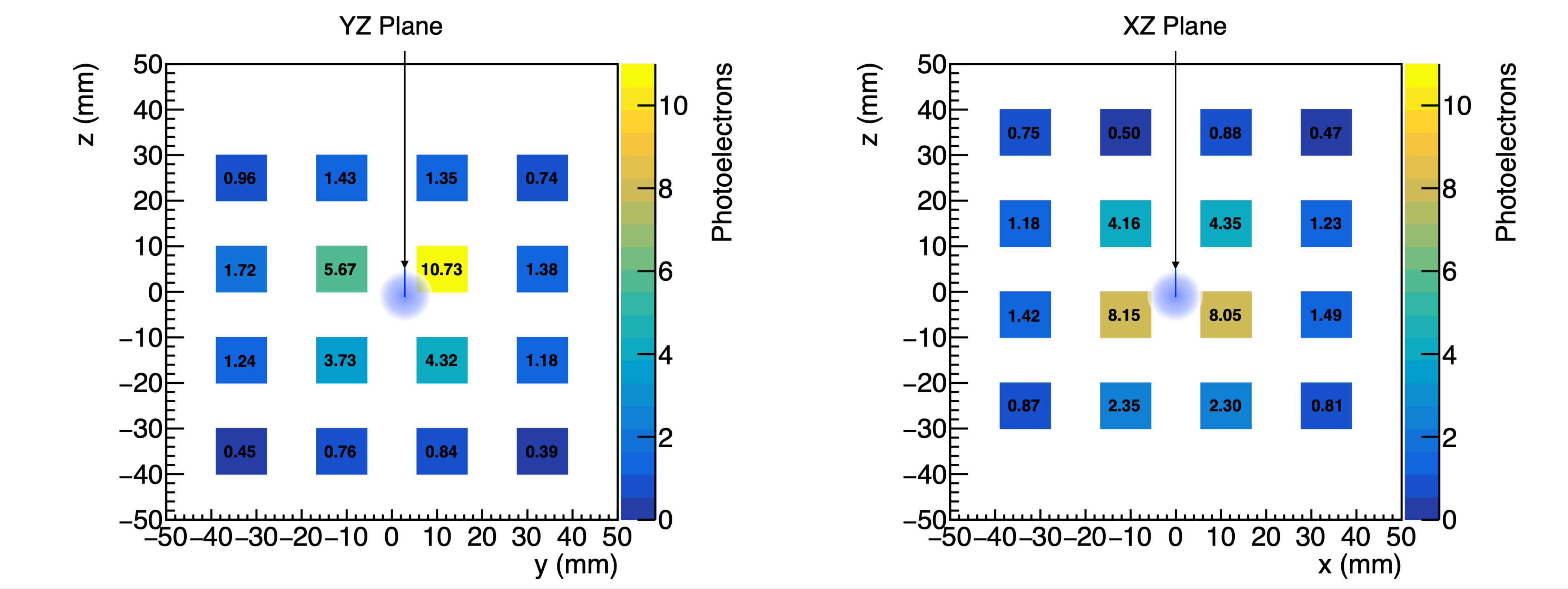}
    \caption{The experimental data from Fig.~\ref{fig:parameter_search_results} shown overlaid on a detector sketch.  The tip of the arrow shows the terminus position of the laser fiber.  The light ball is located approximately one mean free path directly below the laser fiber tip.}
    \label{fig:average_event_laser}
\end{figure*}

The optimal parameter combination was found with a genetic algorithm, using the Multi-variate Analysis (TMVA) package in ROOT.  In a genetic algorithm, different combinations of parameters represent individuals in a population.  Each individual can be assessed for fitness, and the average fitness of a population can be made to increase over many generations.  For each individual in a generation, a Geant4 simulation was run with the corresponding parameter values.  The output of the simulation was the number of photoelectrons per channel per pulse, just as in the experiment.  The fitness of an individual was calculated as the value of $\chi^2$ between the simulated and experimental data.  A comparison of the experimental and simulated signal distributions is shown in Fig.~\ref{fig:parameter_search_results}, and the set of optimized parameters is shown in Table~\ref{table:GA_parameters}.  The experimental data overlaid onto the two detector planes are shown in Fig.~\ref{fig:average_event_laser}.  The number of photoelectrons detected in each channel per pulse depends on the proximity of the WLS fiber to the laser-coupled fiber.  Since the light confinement is stochastic, it is possible for photons to propagate to more distant fibers.  This is apparent by the flat yield of $<1$~photons/pulse for fibers in the corners of the detector shown in Fig.~\ref{fig:average_event_laser}.

To investigate the compatibility of the experimental and simulated distributions with the optimized parameters, a $\chi^2$ test was applied to each laser pulse measurement, with the simulated distribution as a reference.  The mean $\chi^2$/NDF value was $0.6\pm0.2$ and the mean p-value was $0.9\pm0.1$, where the uncertainties are the standard deviations of the distributions.   
\begin{table}[ht]
\centering
 \begin{tabular}{p{0.7in} l l}
 \hline \hline
    Parameter  & Expected Value & Optimized Value \\
    \hline
    $x$  & $0\pm5$~mm  &  -0.09~mm\\
    $y$  & $0\pm5$~mm & 2.62~mm \\
    $z$  & $0\pm5$~mm & 4.42~mm \\
    Reduced scattering length & $5\pm3$~mm & 5.7 mm \\
    Absorption length & $3\pm3$~m & 0.169 m\\
    Reflectivity & $0.3\pm0.1$ & 0.48 \\
    \hline \hline
 \end{tabular}
\caption{The results of the parameter optimization for oWbLS2.  The parameters were optimized by varying the optical parameters in a simulation, comparing the results with the experimental dataset, and selecting the parameter values that minimized the residual between them.  The parameter space was searched with a genetic algorithm.  The experimental dataset was generated by pulsed light from a 407~nm fiber-coupled laser near the center of the detector.}
\label{table:GA_parameters}
\end{table}

The absorption length of DIN is known to be shorter than LAB, but DIN could nevertheless be desirable for its higher light yield.  The measured absorption lengths of pure DIN at 430~nm range from 1~m~\cite{Bonhomme2022SafeDetectors} to 4.2~m~\cite{Song2013FeasibilityScintillator} and are expected to be lower at shorter wavelengths.  The value of 0.169~m found here is therefore lower than expected.  There are several possible explanations for this discrepancy.  First, the light attenuation during transport in WLS fibers may be greater than modeled.  While the model of light transport in the fibers used here was previously validated~\cite{Wilhelm2023EvaluationFibers}, there are several differences in this experimental campaign from prior experiments.  One is that the fibers were occasionally exposed to ambient fluorescent room lighting for about a year, and may have degraded.  A dedicated study on the effect of ambient light on WLS fibers has not been carried out, but Ref.~\cite{Chung1993EffectsFibers} found that a permanent 38\% reduction of the attenuation length in a single-clad Kuraray scintillating fiber could be induced by a 137-hour exposure to ambient light.  Kuraray estimates the attenuation length of Y-11 fiber to be 3.5~m~\cite{2022KurarayDatasheet}, which becomes 2.2~m after a 38\% degradation.  Assuming a 0.5~m path for a photon to reach the MAPMT, this would result in an 8\% signal loss in our detector.  Secondly, while the fiber bending radius is within the manufacturer's recommendation, the bending necessary to route the fibers to the MAPMT may nevertheless introduce unaccounted losses.  The oWbLS is not pure DIN and may have a shorter absorption length due to the addition of the surfactant.  Furthermore, the optical properties of the oWbLS may have degraded over time due to exposure to ambient oxygen and light.  While the absorption length calculated here primarily served to enable simulations for vertex reconstruction in Sec.~\ref{sec:reconstruction}, it may not accurately reflect the true optical properties of the material. Additional experiments, described in Sec.~\ref{sec:Future_Work}, are planned to directly measure the optical properties of oWbLS and address these discrepancies.  

\subsection{Gamma ray response}
\begin{figure*}
    \centering
    \includegraphics[width=\textwidth]{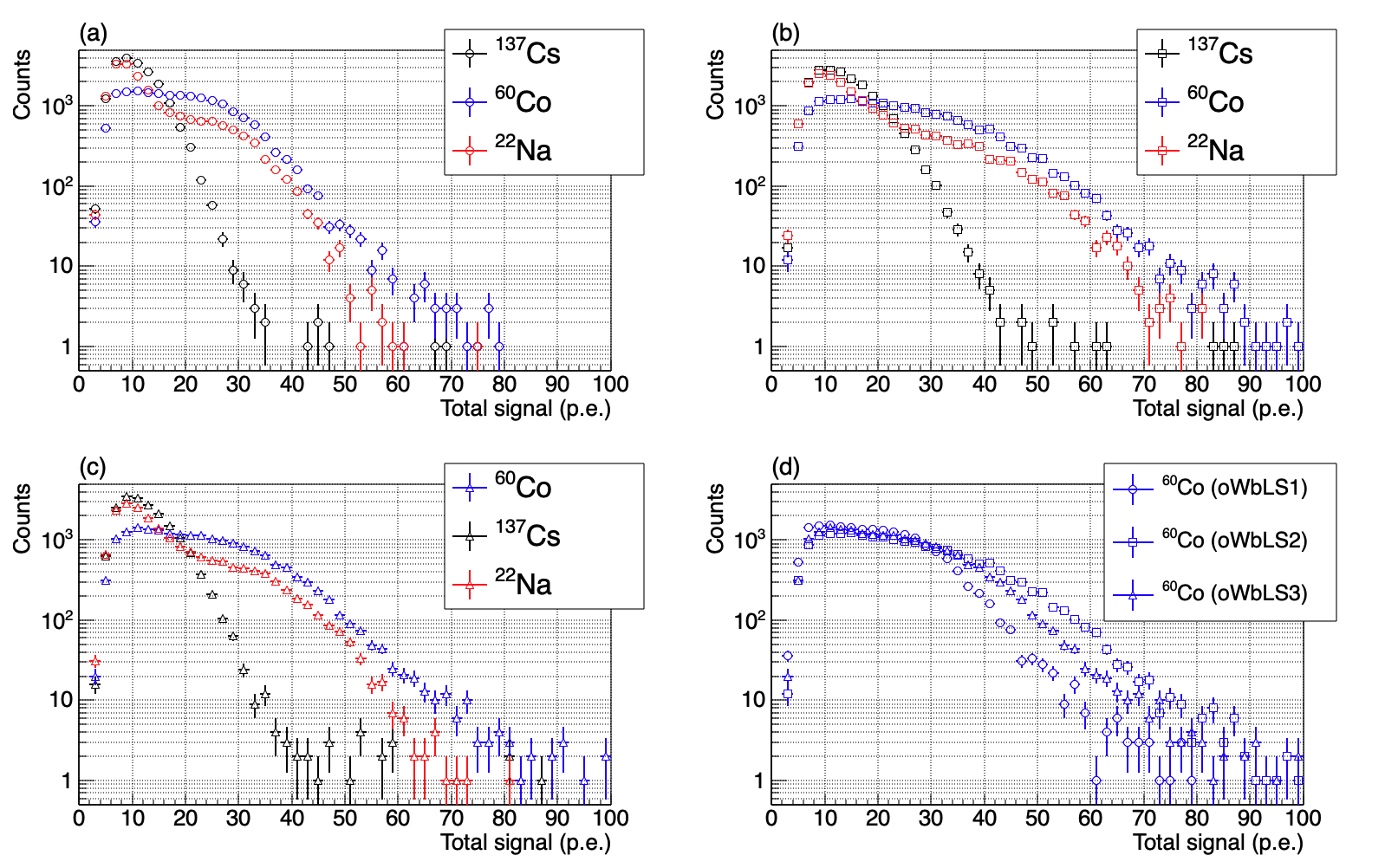}
    \caption{Response to gamma rays from three radioisotopes for (a)~oWbLS1, (b)~oWbLS2, and (c)~oWbLS3.  Among these, oWbLS1 had the longest reduced scattering length, leading to the sharpest Compton features, while oWbLS2, with the shortest reduced scattering length, provided optimal light collection.  (d)~Response of the three formulations of oWbLS to gamma rays from $^{60}$Co for ease of comparison.  The light collection efficiency increases as the reduced scattering length decreases.  The 511~keV gamma ray from $^{22}$Na has the lowest energy of any from the radioisotopes shown.  The maximum energy deposited by a single Compton scatter of a 511~keV gamma ray is approximately 340~keV, so the detector is sensitive to energy depositions of events with energy at least this low.}
    \label{fig:gamma_response}
\end{figure*}

To assess the response of the detector to ionizing radiation, gamma-ray sources were used.  In all cases, the gamma-ray sources were low activity ($<5~\mu$Ci), and were placed directly on top of the lid of the detector body, centered in the horizontal plane.  To a good approximation, the energy deposited in the detector by electrons is linearly proportional to the amount of emitted light.  In scintillators, it is generally desirable for the signal measured to be proportional to the amount of light created.  However, the amount of light collected in a channel also strongly depends on the event proximity to the fiber corresponding to that channel.  The summed signal from all channels carries the information on the total energy deposited; however, that information is convolved with the position-dependent light collection efficiency.  Figure~\ref{fig:gamma_response} shows the integral detector response for gamma rays from three radioisotopes, for each of the formulations of oWbLS.  The general trend is that increased scattering in the liquid improves light collection as it increases the probability of a photon reaching the fiber.  The spectral shapes also differ among various scintillators due to their different optical properties.  oWbLS1, the most transparent of the formulations, has well-defined Compton features, whereas the more highly scattering formulations (like oWbLS2), have Compton features that are broadened.  This is because the dependence of collection efficiency on event distance to a fiber increases as the reduced scattering length decreases.
\subsection{Comparison to simulation}
\begin{figure*}
    \centering
    \includegraphics[width=\textwidth]{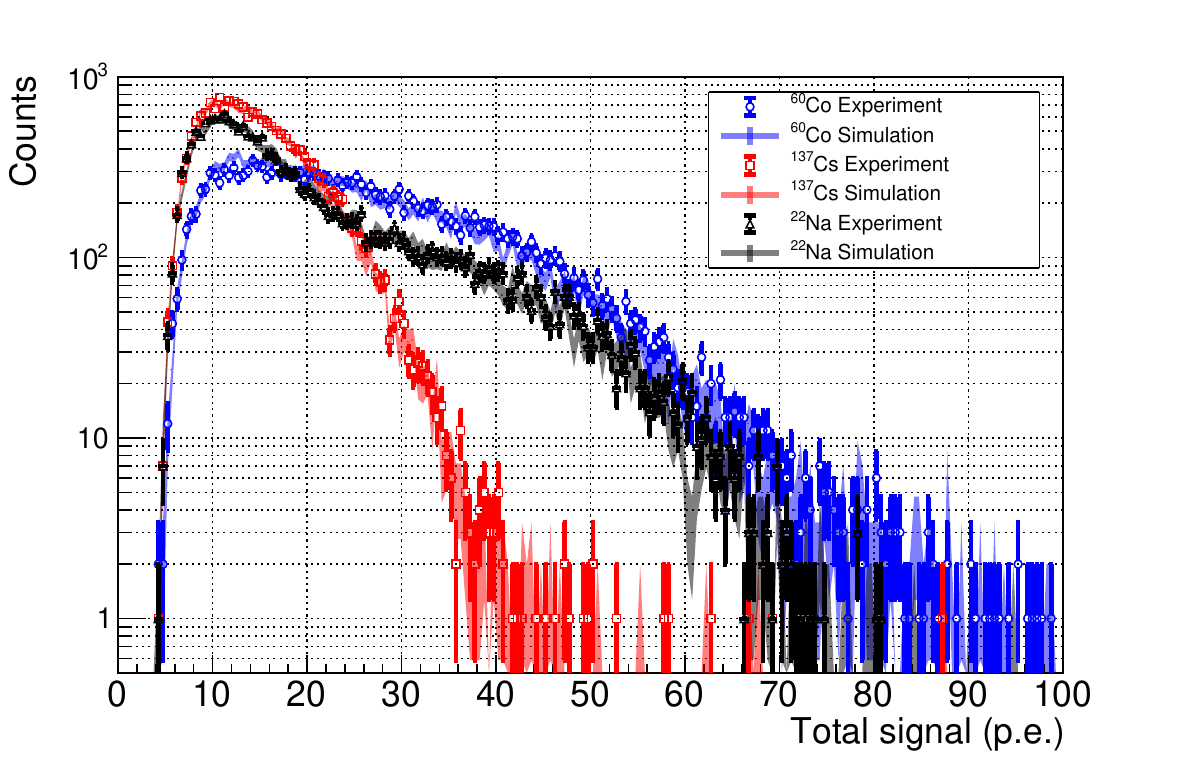}
    \caption{The comparison of experimental and simulated detector response for gamma rays from three radioisotopes.  The simulation results are shown as a shaded region.  The simulation parameters are those described in Table~\ref{table:GA_parameters}.  More details about the simulation methodology are available in Ref.~\cite{Wilhelm2023EvaluationFibers}.}
    \label{fig:sim_comp}
\end{figure*}

To further validate the model of oWbLS2, we simulated the detector response to gamma-ray sources in the configuration of the experiment described above.  The experimental and simulated gamma-ray response for the detector filled with oWbLS2 is shown in Fig.~\ref{fig:sim_comp}.  The parameters used in the simulation were based on those found in Sec.~\ref{sec:oWbLS_optical_parameter_determination}.  The absorption length was extrapolated from the measurement at 407~nm to the range of wavelengths 345-455~nm using the shape of the DIN absorption spectrum in Ref.~\cite{Song2013FeasibilityScintillator}.  The value of the reduced scattering length measured at 407~nm was extrapolated to the 342--457~nm range using the $1/\lambda^4$ dependence of the Rayleigh scattering cross-section~\cite{Rayleigh1899OnSky}.  The $\chi^2/\textnormal{NDF}$ values between the experimental and simulated spectra are 1.45 for $^{60}$Co, 1.13 for $^{137}$Cs, and 1.23 for $^{22}$Na.
\subsection{Topological reconstruction}
\label{sec:reconstruction}
We investigated the use of channel-by-channel signals for an event to reconstruct event topology.  The reconstructions were based on the data from the laser-pulse injection experiments to obtain ``ground-truth'' values for vertex position and event intensity.  The simplest method to reconstruct the event position is to calculate a quantity analogous to the center of mass (CoM) for all channels.  For instance, in the $z$ direction:
\begin{align}
    \textnormal{CoM}_z=\frac{\sum_{i=0}^{32}s_ip_{i,z}}{\sum_{j=0}^{32}s_j},
\end{align}
where $s_j$ is the signal in the $j$th fiber, and $p_{j,z}$ is the position of the $j$th fiber in the $z$ direction.  The center of the detector is taken as the origin.  As noted in Sec.~\ref{sec:detector_design}, while all 32 channels give information about the event position in the $z$ direction, only 16 provide information in the $x$ and $y$ directions.

This method has several shortcomings.  First, it implicitly assumes that light is created isotropically, which is not the case with this laser injection calibration experiment. 
Secondly, it does not inherently allow for multiple light centers (though in principle it could be extended to account for \textit{e.g.} two Compton scatters if the light balls from the two events did not overlap in any dimension and CoM calculations could be done for each region of interest). 

\begin{figure*}
    \centering
    \includegraphics[width=\textwidth]{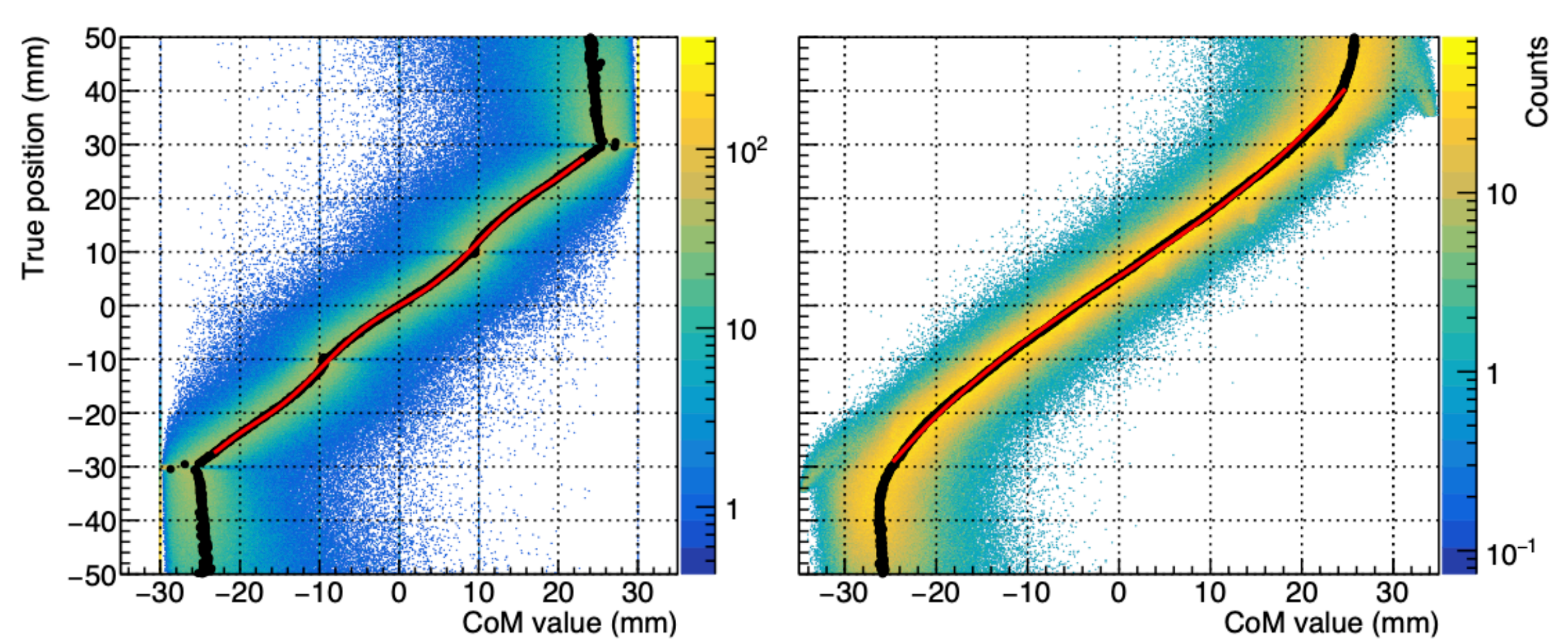}
    \caption{Mapping of CoM and true positions of events for the (left) $x$ and $y$ dimensions and (right) the $z$ dimension.  The histogram shows the true and reconstructed positions for $3\times10^6$ simulated events that varied in the range of 5000--30000 photons/pulse.  The black dots show the centroids of Gaussian distributions fit to slices of the histogram.  The red line shows the polynomial fit (9$^\textnormal{th}$ order for $x$ and $y$, and 3$^\textnormal{rd}$ order for $z$) to the Gaussian centroids.}
    \label{fig:CoM_map}
\end{figure*}

The first shortcoming can be partially addressed by systematic correction, \textit{i.e.}, mapping the true source positions in each dimension as a function of the solved CoM positions.  This relationship was calculated by simulating $3\times10^6$ events and creating 2D histograms of the true and apparent positions for each dimension.  Then, for each slice of vertical bins (representing a small range of true positions along an axis) a Gaussian distribution was fit to the resulting CoM positions.  Lastly, a polynomial (9$^\textnormal{th}$ order for $x$ and $y$, and 3$^\textnormal{rd}$ order for $z$) function was fit to the Gaussian centroids, and this function was used to transform experimental CoM values to true values.  The histograms and fits are shown in Fig.~\ref{fig:CoM_map}, illuminating an additional shortcoming of the CoM reconstruction approach: events that take place near the edges of the detector do not result in unique CoM values.  This reduces the fraction of the detector volume for which this technique is applicable.  The ranges of suitable values (shown as the range of the red fit lines in the plot) are approximately $-27$ to 27~mm in $x$ and $y$, and $-30$ to 40~mm in $z$.  This fiducial region corresponds to a volume of 0.20~L, or about 20\% of the overall detector volume.

\begin{figure}
     \centering
     \includegraphics[width=\textwidth]{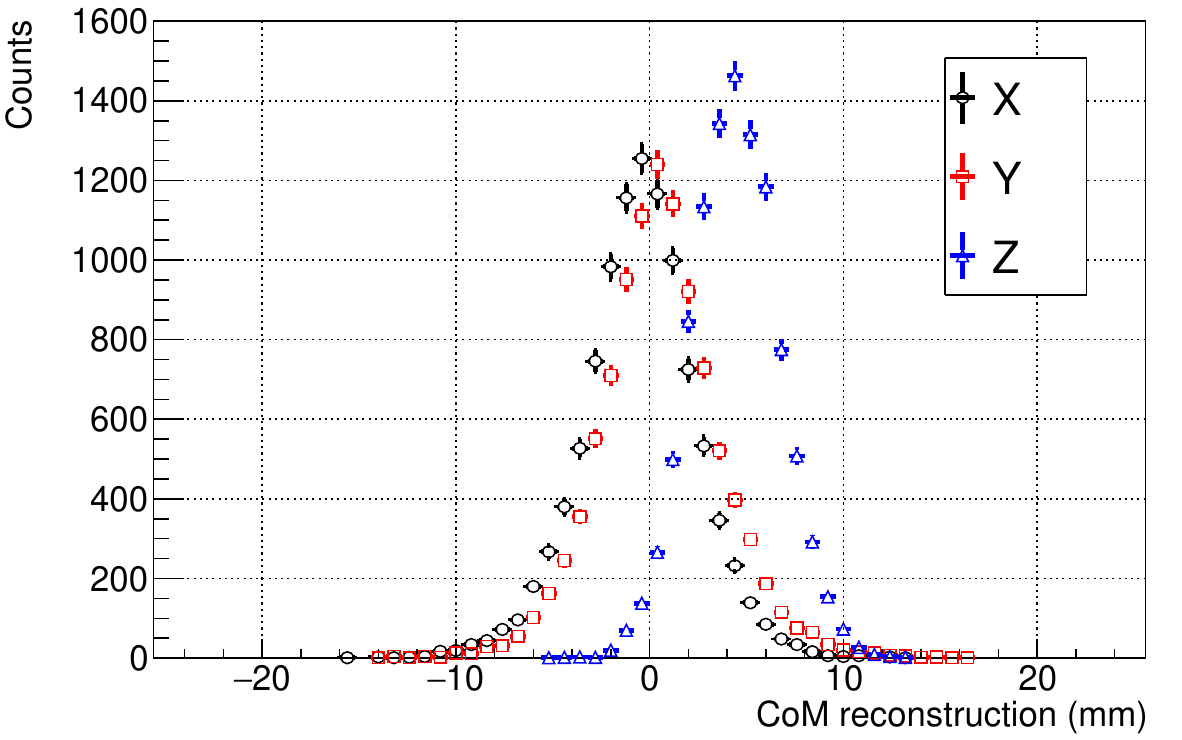}
     \caption{Distribution of position reconstruction errors from the CoM method in each dimension for pulses of approximately 20000 photons/pulse at a nominal position of $(0,0,0)$.  The standard deviations are 2.82~mm, 2.96~mm, and 2.22~mm for $x$, $y$, and $z$, respectively.  As expected, the distributions of errors in $x$ and $y$ have similar widths, and the distribution of errors in $z$ is narrower due to the increased fiber density in this direction.}
     \label{fig:CoMs}
\end{figure}

Figure~\ref{fig:CoMs} shows the results of the CoM reconstruction approach for the 10000 laser pulses obtained from the experiment shown in Fig.~\ref{fig:laser_measurement}.  The number of photons injected corresponds to an event depositing approximately 1.6~MeV in the detector.  The position of the reconstructed values are $x=-0.48\pm0.03$~mm, $y=0.48\pm0.03$~mm, and $z=4.43\pm0.02$~mm, where the uncertainties are the errors of the centroid values of Gaussian fits.  These shifts from the (0,0,0) position arise from a combination of the offset of the true initial position of photon injection and variations in the responses of each channel.  

\begin{figure}
     \centering
     \includegraphics[width=\textwidth]{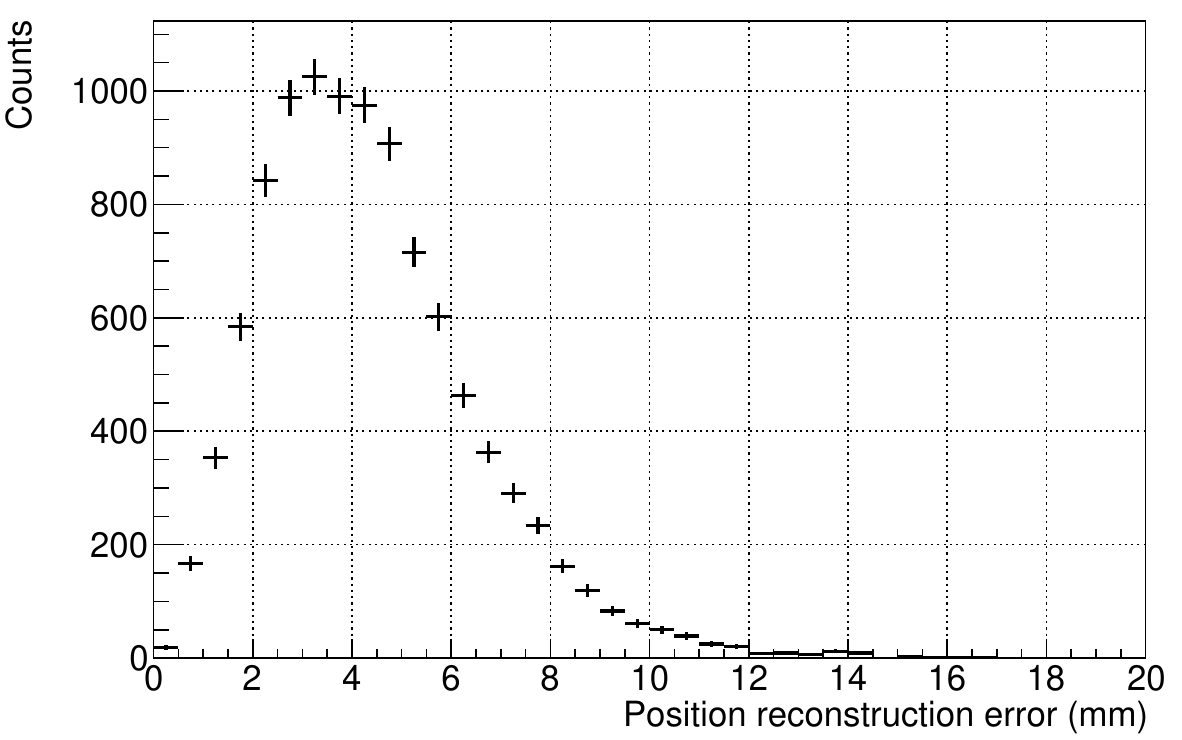}
     \caption{Distribution of position reconstruction errors using the CoM method for a point-like event comprising approximately 20000 initial photons near the center of the detector, with a mean of 4.35~mm and a standard deviation of 2.15~mm.}
     \label{fig:total_error}
\end{figure}

\begin{figure}
    \centering
    \includegraphics[width=\textwidth]{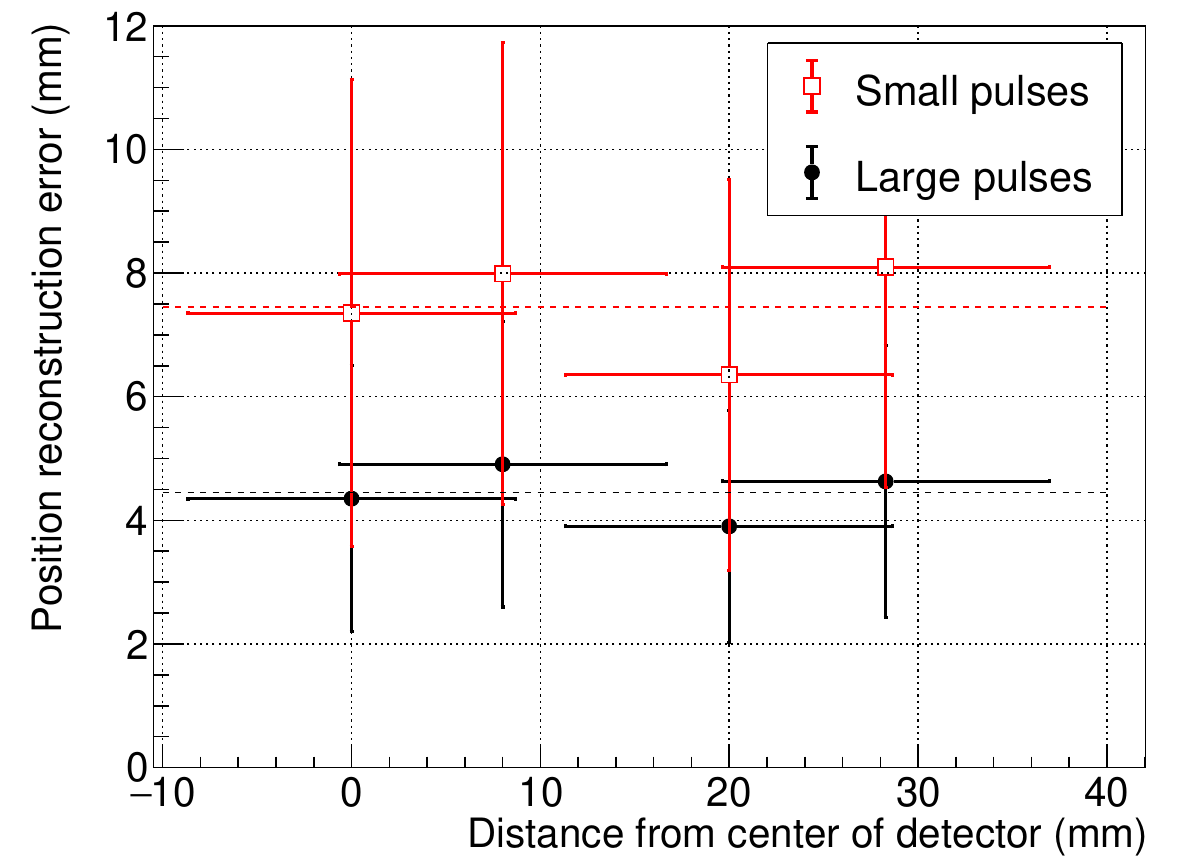}
    \caption{Position reconstruction errors for the detector positions listed in Table~\ref{table:reconstruction_positions}.  The large pulses had an intensity of $20000\pm2000$ photons/pulse, and the small pulses were $10000\pm1000$ photons/pulse.  The average reconstruction error (shown as dotted horizontal lines) for the large pulses was 4.4~mm and for the small pulses was 7.4~mm.  The horizontal error bars represent the uncertainty in offset from the center of the detector, and the vertical error bars show the standard deviation of the reconstruction position errors.}
    \label{fig:CoM_errors}
\end{figure}

The standard deviations for the three distributions are $x$:~2.82~mm, $y$:~2.96~mm, and $z$:~2.22~mm.  The superior performance in the $z$ direction is due to the increased fiber density and indicates the potential precision of a detector with a fiber pitch of 1~cm.  While the distributions of the CoM errors have a Gaussian shape, the transformation from CoM to true positions using the mapping function shown in Fig.~\ref{fig:CoM_map} skews the distributions.  The distribution of total position reconstruction error is found by shifting the individual dimensional distributions such that their mean value vanishes (to remove bias) and calculating the Euclidean distance for each event from the origin.  The distribution of total errors for the dataset shown in Fig.~\ref{fig:CoMs} is shown in Fig.~\ref{fig:total_error}.  The distribution has a mean value of 4.35~mm and a standard deviation of 2.15~mm.  
This experiment was repeated for four positions in the detector with two pulse intensities.  The large pulses had an intensity of $20000\pm2000$~photons/pulse, which correlates to approximately 1.6~MeV of energy deposition in oWbLS.  The small pulses had an intensity of $10000\pm1000$~photons/pulse, which is equivalent to 0.8~MeV of energy deposited in the detector.  The four positions, along with their nominal offsets from the center of the detector, are listed in Table~\ref{table:reconstruction_positions}.  The results of CoM analysis are shown in Fig.~\ref{fig:CoM_errors}.  The position reconstruction error is independent of the laser injection position within the fiducial region, indicating a homogeneous performance.  The reconstruction error of the $y$-direction was higher than the error of the $x$-direction for the measurement taken at $(0,-8,0)$~mm, indicating that there may be small inhomogeneities within individual cells in the fiducial region.  This effect appears to be minor, as the overall reconstruction error at this position was similar to the others, but should be studied in more detail in future prototypes.
 \begin{table}[ht]
\centering
 \begin{tabular}{c | c | c | c | p{0.4in}}
 \hline \hline
    Position & $x$ (mm) & $y$ (mm) & $z$ (mm) & Nominal offset (mm) \\
    \hline
    1 & $0\pm5$   & $0\pm5$   & $0\pm5$ & $0\pm9$\\
    2 & $0\pm5$   & $-8\pm5$  & $0\pm5$ & $8\pm9$\\
    3 & $0\pm5$   & $-20\pm5$ & $0\pm5$ & $20\pm9$\\
    4 & $20\pm5$  & $-20\pm5$ & $0\pm5$ & $28\pm9$\\
    \hline \hline
 \end{tabular}
\caption{Positions used for light injection experiments to assess variations in reconstruction precision.  Nominal offset refers to the distance between the laser and the center of the detector volume.}
\label{table:reconstruction_positions}
\end{table}
\subsection{Energy Resolution}
\label{sec:enery_res}
\begin{figure*}
    \centering
    \includegraphics[width=\textwidth]{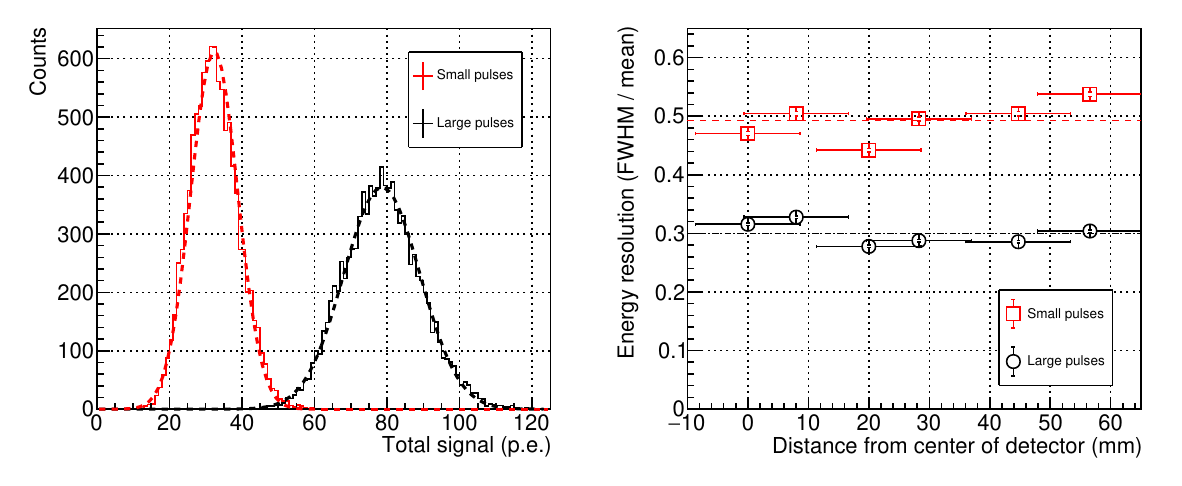}
    \caption{(Left) Distribution of p.e. per pulse for light injected into the center of the detector.  Each distribution is fit with a Gaussian, shown as a dotted line.  (Right) Energy resolution as a function of distance from the center of the detector.  The energy resolution values are calculated from the Gaussian fitting of the individual total signal distributions.  The mean energy resolution values for large ($20000\pm2000$ photons/pulse) and small ($10000\pm1000$ photons/pulse) pulses are shown as a dotted line.}
    \label{fig:energy_res}
\end{figure*}

The position-dependent energy resolution was calculated from the same datasets that were used for the CoM analysis.  In addition, two positions outside the fiducial region were also included.  These positions were $(20,-40,0)$~mm and $(-40,-40,0)$~mm.  To determine the energy resolution, the total signal spectra for pulses of two different heights at each position were fit to a Gaussian shape, with mean $\mu$ and standard deviation $\sigma$.  The energy resolution is then
\begin{align}
    \frac{\Delta E}{E} = \frac{\textnormal{FWHM}}{\mu}, 
\end{align}
where
\begin{align}
    \textnormal{FWHM} = 2.35 \sigma
\end{align}
is the full width at half maximum of the Gaussian fit.  An example of the total signal spectra for one position and energy resolutions for all six positions is shown in Fig.~\ref{fig:energy_res}. Similarly to the position reconstruction, the energy resolution is independent of the laser injection position, again indicating a homogeneous response. The energy resolution for large pulses (corresponding to 1.6~MeV) was $30\pm2$~\%, whereas for the small pulses (corresponding to 0.8~MeV) was $49\pm3$~\%. The uncertainties are the standard deviations of the set of energy resolutions reconstructed from all positions.  

This measurement does not account for the intrinsic energy resolution of the scintillation liquid, which arises from the statistical fluctuation in the number of scintillation photons produced at a given energy deposition by a charged particle.  The intrinsic energy resolutions for various scintillation liquids based on LAB and pseudocumene were found to vary in the range of 1.3--3.3\%~\cite{Smirnov2023NoteScintillators}.  While a similar measurement has not been done for oWbLS, the contribution of oWbLS to the obtained energy resolution is likely in the same range.  Using laser pulses to determine energy resolution also introduces broadening as a result in the variance in intensity of the laser pulses  The laser used in this experiment has a pulse-to-pulse stability of $<5$~\% \cite{BorisKhouryNKTPhotonics2024PrivateCorrespondence}, which partially offsets the error related to the intrinsic energy resolution of oWbLS.  

There are opportunities to dramatically improve this energy resolution which will be incorporated in future designs, including increasing the absorption length of the bulk scintillator and substituting high efficiency silicon photomultipliers (SiPMs) for the relatively low quantum efficiency MAPMT used in this prototype.
\section{Conclusion}
\label{sec:Conclusion}
A 32-fiber, 1~L organic gamma-ray detector capable of spectroscopy and vertex reconstruction for point-like events was demonstrated.  The position reconstruction precision was 4.4~mm for events of an intensity corresponding to 1.6~MeV and 7.4~mm for events of 0.8~MeV, within a centralized fiducial region of the detector. 

 \subsection{Discussion}
 A useful comparison with this performance is to a hypothetical detector volume with cubical, physically segmented voxels.  If the events are homogeneously distributed and all reconstruct to the center of the voxel in which they took place, the mean reconstruction error is
\begin{align}
    \bar{e} = \frac{1}{L^3}\int^{L/2}_{-L/2}\int^{L/2}_{-L/2}\int^{L/2}_{-L/2}\sqrt{x^2+y^2+z^2}\,dx\,dy\,dz , 
\end{align}
which evaluates to approximately $0.48L$, where $L$ is the side length of the cubical voxel.  The number of voxels required to achieve a mean position reconstruction error of 4.4~mm can therefore be calculated by
\begin{align}
    L = 4.4\textnormal{ mm}/0.48 = 9.17\textnormal{ mm}.
\end{align}
Segmenting the fiducial region of the cube with voxels of this size would require approximately 265 voxels.  The 7.4~mm position reconstruction error obtained with events corresponding to 0.8~MeV could be achieved with approximately 56 voxels, or about double the number of channels used in this prototype.  It should be explicitly noted that the position resolution achieved by a physically segmented detector does not depend upon the number of photons produced by an event (above some minimum threshold), whereas it does in an opaque scintillator.  Moreover, this technique allows an enormous advantage when scaling up the detector volume.  For example, adding another ring of fibers, (\textit{i.e.}, a $6\times6$ grid as opposed to the current $4\times4$ per side) to create a fiducial volume of approximately 1~L would require an additional 40 channels, but approximately another 1000 physically segmented voxels of side length 9.17~mm.  This analysis assumes a single, point-like interaction in the detector.  Physically segmented detectors may be superior for distinguishing, for instance, two simultaneous events that take place near each other, but with sufficient separation to be located in two voxels.  Additional studies are needed to determine the minimum separation between events that can be resolved in an opaque scintillation detector.    

3D reconstruction has also been demonstrated with a detector that employs 2D voxels (rods).  The rods are rectangular columns of transparent plastic scintillator that are optically separated from each other.  Position reconstruction in the third dimension may be accomplished by comparing the amount of light collected or the relative time of arrival of scintillation photons~\cite{Sutanto2021SANDD:Scintillator} at the two ends of a rod.  Since the readout array is laid out in two planes, the number of channels scales similarly with volume to the detector described in this study.  Furthermore, the energy resolution of such a system can reach 18-22\% at the $^{137}$Cs Compton edge, which is superior to the results shown in Sec.~\ref{sec:enery_res}.  However, there are two disadvantages of the rod technique compared to an opaque scintillator.  First, sub-voxel position resolution is generally not achievable in the dimensions orthogonal to the rods, whereas sub-fiber pitch resolution has been demonstrated here for opaque scintillators. And secondly, fast timing electronics are necessary to reconstruct the position along the rods.  While timing information may be useful in other opaque scintillator designs, it was not used here beyond the correlation of pulses into events.
\subsection{Future work}
\label{sec:Future_Work}
There are many opportunities to extend this work, which can be grouped into three categories.  

The first category relates to the optimization of detector design.
This prototype detector was not optimized for a particular application by adjusting the scintillator composition and doping.  The quality of topological reconstruction could almost certainly be improved by tuning the positioning and pitch of WLS fibers. 
Additionally, using SiPMs instead of a single MAPMT could yield at least two benefits: a higher quantum efficiency and the ability to place photosensors close to the edge of detector volume, which would reduce the length and bending of the WLS fibers and, in turn, reduce light loss.

The second category relates to opaque scintillator media. An apparatus to optically characterize opaque liquid scintillators should be developed, potentially by imaging the diffuse reflectance pattern on the surface of the liquid in response to an incident pencil beam of light, as is done in Refs.~\cite{Kienle1997ImprovedMedium,Juttula2014InstrumentAngle}.  Many media can be investigated for opaque scintillation detector development.  For example, plastics and pressed organic polycrystals can be made to be highly scattering~\cite{Polupan2023PeculiaritiesPropagation}.  Additionally, the reduction in transparency requirements for detectors that use this technology raises exciting possibilities for detector doping at unprecedented levels with, for example, nanoparticles that include material with a high atomic number to increase both light scattering and the photoelectric absorption for gamma-ray detection~\cite{Taylor2011NanofluidCollectors, Mutreja2023HighAspects}. 

The third category of improvements relates to improving particle identification and topological reconstruction performance.
Machine learning for event topological reconstruction could be appropriate for this application.  One such method is to train a neural network by simulated data to generate maximum likelihood estimations of event characteristics given a series of observables~\cite{Eller2023APrinciples} (in this case, the number of photoelectrons in each channel).  
Lastly, the base solvent used in oWbLS (DIN)~\cite{Paweczak2013StudiesMethod} as well as transparent water-based liquid scintillators~\cite{Ford2022Pulse-shapeScintillators} have been shown to have the ability to discriminate between neutron and gamma-ray interactions based on the difference in the resulting pulse shape.  The feasibility of this approach in the opaque regime should be investigated.

\subsection{Acknowledgements}

We are thankful to the CNPq/CAPES in Brazil, the McDonald Institute providing FVRF support in Canada, the Charles University in the Czech Republic, the CNRS/IN2P3 in France, the INFN in Italy, the Fundação para a Ciência e a Tecnologia (FCT) in Portugal, the CIEMAT in Spain, the STFC/UKRI/Royal Society in the UK, the University of California at Irvine, Department of Defense, Defense Threat Reduction Agency (HDTRA1-20-2-0002) and the Department of Energy, National Nuclear Security Administration, Consortium for Monitoring, Technology, and Verification (DE-NA0003920), Brookhaven National Laboratory supported by the U.S. Department of Energy under contract DE-AC02-98CH10886, U.S. National Science Foundation-Major Research Instrumentation Program, Deep Learning for Statistics, Astrophysics, Geoscience, Engineering, Meteorology and Atmospheric Science, Physical Sciences and Psychology (DL-SAGEMAPP) at the Institute for Computational and Data Sciences (ICDS) at the Pennsylvania State University in the USA for their provision of personnel and resources.

%
\printbibliography
\end{document}